\documentclass[a4paper,11pt]{article}
\pdfoutput=1 

\usepackage{jheppub} 

\usepackage[T1]{fontenc} 

\usepackage{amsmath,amssymb,epsfig,ulem,color,slashed}
\allowdisplaybreaks  

\def \beq{\begin{equation}}
\def \eeq{\end{equation}}
\def \bea{\begin{eqnarray}}
\def \eea{\end{eqnarray}}
\def \Op{O}
\def \Wc{C}

\title{\boldmath Constraints on CP-violating Higgs couplings \\ to the third generation}

\author[a]{Joachim Brod,}
\author[b]{Ulrich Haisch}
\author[a]{and Jure Zupan}

\affiliation[a]{Department of Physics, University of Cincinnati, \\ Cincinnati, Ohio 45221,USA}
\affiliation[b]{Rudolf Peierls Centre for Theoretical Physics,
    University of Oxford, \\ OX1 3PN Oxford, United Kingdom}

\emailAdd{joachim.brod@uc.edu}
\emailAdd{u.haisch1@physics.ox.ac.uk}
\emailAdd{zupanje@ucmail.uc.edu}

\abstract{Discovering CP-violating effects in the Higgs sector would
  constitute an indisputable sign of physics beyond the Standard
  Model. We derive constraints on the CP-violating Higgs-boson couplings 
  to top and bottom quarks as well as to tau leptons from low-energy 
  bounds on electric dipole moments, resumming large logarithms when 
  necessary. The present and future projections of the sensitivities and 
  comparisons with the LHC constraints are provided. Non-trivial constraints 
  are possible in the future, even if the Higgs boson only couples to the 
  third-generation fermions.}
  
\preprint{NSF-KITP-13-229}

\begin{document} 

\maketitle

\flushbottom

\section{Introduction}
\label{sec:introduction}

There is steady experimental progress in measuring the Higgs-boson
couplings. Assuming for simplicity that deviations from the Standard
Model (SM) manifest themselves predominantly in a single coupling, the
couplings of the Higgs to $Z$ and $W$ bosons are known with an
uncertainty of ${\mathcal O}(20-30\%)$, and to the third-generation
fermions $t$, $b$, and $\tau$ with ${\mathcal O}(30\%)$, ${\mathcal
  O}(40\%)$, and ${\mathcal O}(60\%)$ relative errors, respectively
(the sensitivity to the top-quark couplings arises from the loop
processes $gg \to h$ and $h \to \gamma
\gamma$)~\cite{CMS-PAS-HIG-13-005, ATLAS-CONF-2013-034,
  Aad:2013wqa}. The projected sensitivity for the 14~TeV LHC at
300~fb${}^{-1}$ is ${\mathcal O}(4-15\%)$ and ${\mathcal O}(2-10\%)$
at 3000~fb${}^{-1}$ of integrated
luminosity~\cite{Olsen:talk:Seattle}. If deviations from the SM are
found this would suggest that there is new physics~(NP) close to the
TeV scale. In this respect, CP-violating Higgs-boson couplings are
particularly interesting, because any sign of CP violation in Higgs
decays would constitute an indisputable NP signal.

Low-energy probes, such as electric dipole moments~(EDMs), lead to
  severe constraints on CP-violating effects. The purpose of this
paper is to derive the constraints that low-energy measurements set on
CP-violating Higgs couplings to the third generation of fermions. In
complete generality, we can write
\begin{equation} \label{higgsferm}
{\cal L} \supset -\frac{y_f}{\sqrt 2} \left ( \kappa_f \hspace{0.5mm}
\bar f f + i \tilde \kappa_f \hspace{0.5mm} \bar f \gamma_5 f \right )
h \,,
\end{equation}
where $f=t,b,\tau$ and $y_f=\sqrt 2 m_f/v$ is the SM Yukawa coupling
with $m_f$ the fermion mass and $v\simeq246 \, {\rm GeV}$ the
electroweak symmetry breaking vacuum expectation value of the
  Higgs field. The couplings $\tilde \kappa_f$ are CP violating,
while $\kappa_f$ parametrize CP-conserving NP contributions. In the SM
we have $\kappa_f=1$ and $\tilde \kappa_f=0$. Our primary aim is to
derive bounds on the coefficient $\tilde \kappa_f$ using low-energy
data. These can then be used as a useful target for direct searches at
the LHC \cite{Berge:2008wi,Berge:2008dr,Berge:2011ij, Fermilab:CPV, Nishiwaki:2013cma,Htt:CPV}. 
Similarly, one could search for CP-violating Higgs-boson couplings to 
gauge bosons both at the LHC \cite{Godbole:2007cn,Coleppa:2012eh, Stolarski:2012ps, 
Bolognesi:2012mm, Boughezal:2012tz, Chatrchyan:2012jja, ATLAS:spin, 
Freitas:2012kw} or utilizing low-energy observables~\cite{McKeen:2012av}.
Note that there could also be other contributions to the EDMs beyond the ones 
we discuss, for instance from  complex flavor-violating couplings of the Higgs 
 with the corresponding bounds given in \cite{McKeen:2012av,Giudice:2008uua,
 Goudelis:2011un,Blankenburg:2012ex,Harnik:2012pb}.
 
The paper is organized as follows. Focusing first on the CP-violating
Higgs-top couplings we deduce the corresponding constraints from EDMs
in Sec.~\ref{EDM-top} and from the LHC Higgs data in
Sec.~\ref{sec:onshell:Higgs}. The combined effect of the two types of
constraints as well as the projected future sensitivities are
presented in Sec.~\ref{combined-top}. Analogous constraints on
bottom and tau couplings to the Higgs are derived in
Sec.~\ref{EDM:b:tau}. In Sec.~\ref{conclusions} we summarize our main
findings. A series of appendices completes our work. The details about
the renormalization group~(RG) analysis for the neutron EDM are given
in App.~\ref{app:neutronEDM:RG}, while the RG resummation of the
bottom-quark contributions to the neutron EDM is discussed in
App.~\ref{app:lightEDM:RG}.  Finally, in App.~\ref{App:other} the
constraints on the CP-violating couplings of the Higgs to
third-generation fermions arising from flavor-changing neutral current
processes are briefly examined.

\section{Constraints from EDMs}
\label{EDM-top}

EDMs are very sensitive probes of NP that contains new CP-violating
weak phases. They can probe scales as high as $10^8 \, {\rm GeV}$
\cite{Cirigliano:2013lpa,Engel:2013lsa,Pospelov:2005pr}. Here we are
interested in the constraints that the EDM measurements impose on the
CP-violating Higgs-top coupling, i.e.~the coefficient $\tilde
\kappa_t$ in~Eq.~\eqref{higgsferm}. The derivation of constraints on
the Higgs-boson couplings to bottom quarks and tau leptons is
relegated to Sec.~\ref{EDM:b:tau}.

\subsection{EDM of the electron}

\begin{figure}[!t]
\begin{center}
\vspace{-5mm}
\includegraphics[height=0.3 \textwidth]{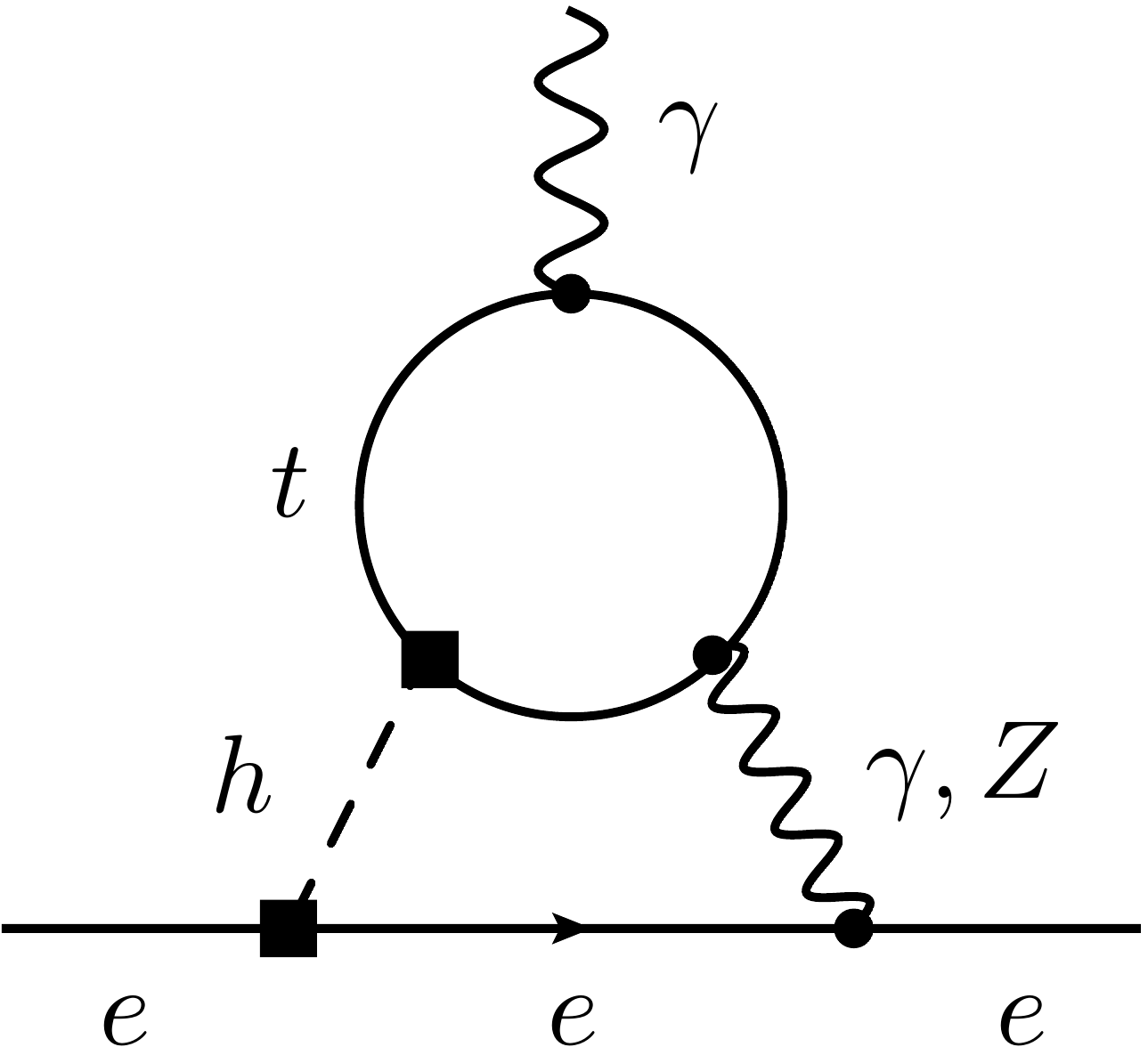} \qquad \qquad \quad 
\raisebox{4mm}{\includegraphics[height=0.28 \textwidth]{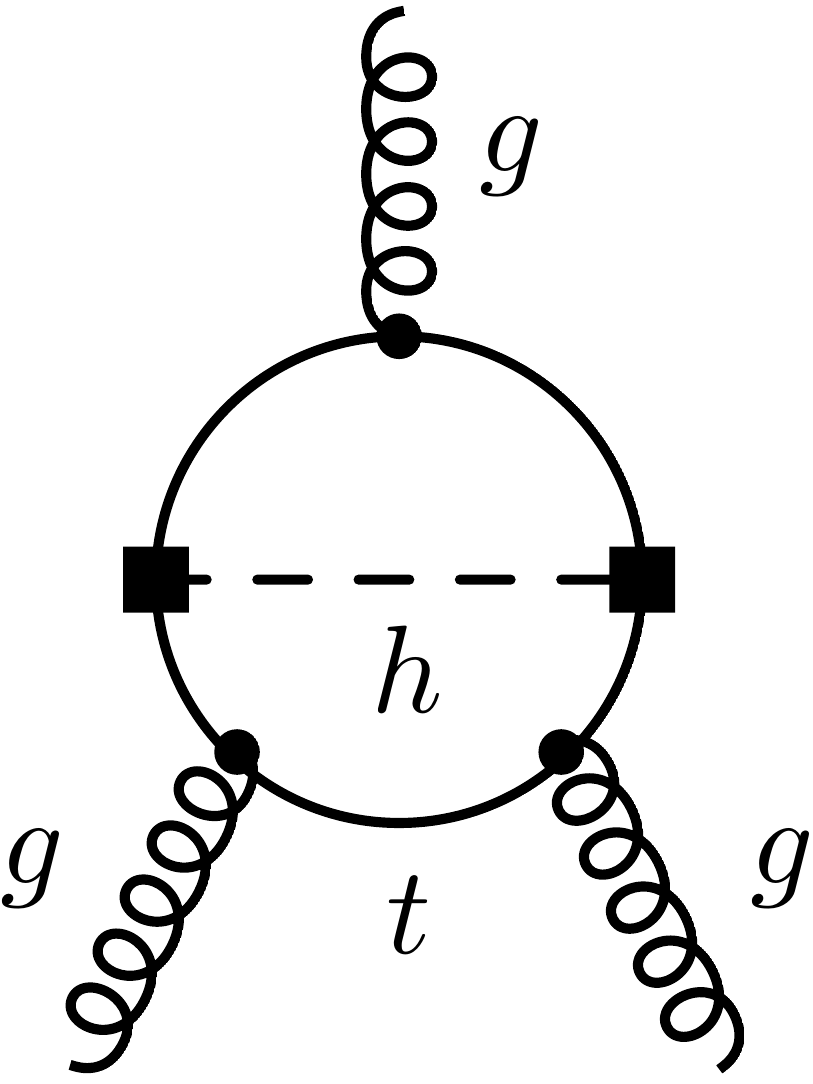} }
\vspace{2mm}
\caption{\label{fig1} Left: Two-loop Barr-Zee contributions to the EDM
  of the electron involving a virtual Higgs boson and a photon or $Z$
  boson.  Right: Two-loop contribution to the Weinberg operator.}
\end{center}
\end{figure}

The CP-violating Higgs-boson coupling to the top quark induces an
electron EDM
\begin{equation}
{\cal L}_{\rm eff} = - d_e \, \frac{i}{2} \, \bar e \hspace{0.5mm}
\sigma^{\mu\nu} \gamma_5  \hspace{0.25mm} e \, F_{\mu\nu}  \,, 
\end{equation}
through a Barr-Zee type two-loop diagram, cf. Fig.~\ref{fig1}
(left). The diagram with the photon propagator gives
\cite{Barr:1990vd}
\begin{equation} \label{first}
\frac{d_e}{e} = \frac{16}{3} \frac{\alpha}{(4\pi)^3} \hspace{0.25mm}
\sqrt{2} G_F  \hspace{0.25mm}  m_e \, \Big[  \kappa_e \tilde \kappa_t
  \, f_1 (x_{t/h}) +  \tilde\kappa_e \kappa_t  \, f_2 (x_{t/h}) \Big
] \,, 
\end{equation} 
where $x_{t/h} \equiv m_t^2/M_h^2$ and the loop functions $f_{1,2}
(x)$ can be written as~\cite{Stockinger:2006zn},\footnote{Note that
  the loop function $f_1(x)$ is real and analytic even for $x>1/4$. In
  particular, in the limit $x\to \infty$, one has $f_1 (x) = \ln x + 2
  + {\cal O} ( 1/\sqrt{x})$.}
\begin{equation}
\begin{split}
f_1 (x) & = \frac{2x}{\sqrt{1-4x}} \left [ {\rm Li}_2  \left ( 1 -
  \frac{1- \sqrt{1-4x}}{2x} \right ) - {\rm Li}_2  \left (  1 -
  \frac{1+ \sqrt{1-4x}}{2x} \right )  \right ] \,, \\[4mm] 
f_2 (x) & = \left (1 - 2 x \right) f_1(x) + 2 x \left (\ln x + 2
\right)   \,. 
\end{split}
\end{equation}
Here ${\rm Li}_2 (x) = -\int_0^x du \, \ln (1-u)/u$ is the usual
dilogarithm.

From Eq.~\eqref{first} it is evident that the electron EDM constraint
on $\tilde \kappa_t$ vanishes in the limit that the Higgs does not
couple to electrons, $\kappa_e, \tilde \kappa_e\to 0$, or by an
appropriate tuning of the ratio $\tilde \kappa_e/\kappa_e$. For
simplicity we will from here on assume that the Higgs coupling to the
electron is CP conserving, so that $\tilde \kappa_e=0$. In this case
the top-quark contribution to the EDM of the electron is (with
{$\alpha \equiv \alpha (0) \simeq 1/137$})
\begin{equation}  
\frac{d_e}{e} =  {3.26} \cdot 10^{-27} \, {\rm cm} \; \kappa_e
\tilde\kappa_t    \hspace{0.5mm} f_1 (x_{t/h})  = {9.0} \cdot 10^{-27}
\, {\rm cm} \; \kappa_e \tilde\kappa_t \,, 
\end{equation}
where in the second equality we used that $f_1 (x_{t/h}) \simeq 2.76$
for $m_t = 163.3 \, {\rm GeV}$~\cite{Alekhin:2012py} and $M_h = 126 \,
{\rm GeV}$.  The 90\% confidence level (CL) limit~\cite{Baron:2013eja}
\begin{equation} \label{eq:bestde}
\left | \frac{d_e}{e} \right | < 8.7 \cdot 10^{-29} \, {\rm cm} \,,
\end{equation}
then translates into
\begin{equation} \label{eq:kappateEDM}
\left | \tilde\kappa_t  \right | < {0.01}\,,
\end{equation}
assuming that the Higgs coupling to the electron is the SM one,
$\kappa_e=1$.

Above we have neglect the two-loop diagram, Fig.~\ref{fig1} (left),
with the $Z$ boson instead of the photon in the loop. Due to
charge-conjugation invariance only the vector couplings of the $Z$
boson enter the Barr-Zee expression for the electron EDM. As a result
the $Z$-boson contribution is strongly suppressed
by~\cite{Barr:1990vd}
\begin{equation}
\left (- \frac{2}{3} \hspace{0.5mm} e^2 \right )^{-1} \, \frac{e^2}{s_W^2 c_W^2} 
\left  (-\frac{1}{4}+s_W^2\right )  \left ( \frac{1}{4} -\frac{2}{3} s_W^2 \right )
\simeq 1.6 \% \,,
\end{equation}
where $s_W^2 \simeq 0.23$ denotes the sine of the weak mixing
angle. Keeping in mind that there is a further suppression by the
$Z$-boson mass, one concludes that the $Z$-boson contribution can be
safely neglected in the phenomenological analysis.

\subsection{EDM of the neutron}

Integrating out the top quark and the Higgs, the CP-violating
Higgs-top coupling Eq.~(\ref{higgsferm}) leads to the following
effective Lagrangian relevant for the neutron EDM
\begin{equation} \label{eq:LeffN}
{\cal L}_{\rm eff} = - d_q \, \frac{i}{2} \, \bar q  \hspace{0.25mm}  \sigma^{\mu\nu} 
\gamma_5 \hspace{0.25mm} q \, F_{\mu\nu} - \tilde d_q \, \frac{ig_s}{2} \, \bar q 
\hspace{0.25mm}   \sigma^{\mu\nu}  T^a \gamma_5 \hspace{0.25mm} q \, 
G_{\mu\nu}^a - w \, \frac{1}{3} \hspace{0.25mm} f^{abc} \, G_{\mu \sigma}^a 
G_{\nu}^{b, \sigma} \widetilde G^{c, \mu \nu} \,,
\end{equation}
where $q=u,d$, while $\widetilde
G^{a,\mu\nu}=\frac12\epsilon^{\mu\nu\alpha\beta}\,G_{\alpha\beta}^a$
is the dual field-strength tensor of QCD, with $\epsilon^{\mu \nu
  \lambda \rho}$ the fully anti-symmetric Levi-Civita tensor
($\epsilon^{0123} = 1$). $T^a$ are the color generators normalized as
${\rm Tr} \hspace{0.5mm} (T^a T^b) = \delta^{ab}/2$. The quark EDM
$d_q$ is obtained from a two-loop diagram similar to
Fig.~\ref{fig1}~(left), but with the electron replaced by a light
quark $q$, while for the chromoelectric dipole moment (CEDM) $\tilde
d_q$ one in addition replaces all photons with gluons. The last term
in the effective Lagrangian \eqref{eq:LeffN} is the purely gluonic
Weinberg operator \cite{Weinberg:1989dx}, which arises from the
two-loop graph in Fig.~\ref{fig1}~(right).

Keeping the dependence on the charge and color factors explicitly, the
two-loop matching at the weak scale $\mu_W= {\cal O} (m_t)$ gives
\begin{equation} \label{eq:dq}
\begin{split}
d_q (\mu_W) &= -4 \hspace{0.25mm}  e  \hspace{0.25mm}
Q_q \hspace{0.25mm} N_c \hspace{0.25mm} Q_t^2 \,
\frac{\alpha}{(4\pi)^3} \hspace{0.25mm} \sqrt{2} G_F  \hspace{0.25mm}
m_q \, \kappa_q  \tilde\kappa_t    \hspace{0.5mm} f_1 (x_{t/h}) \,,\\[2mm] 
\tilde d_q (\mu_W) & = -2  \,
\frac{\alpha_s}{(4\pi)^3} \hspace{0.25mm} \sqrt{2}
G_F  \hspace{0.25mm}  m_q \, \kappa_q \tilde\kappa_t    \hspace{0.5mm}
f_1 (x_{t/h})  \,, 
\end{split}
\end{equation}
for the EDM and CEDM. Here $Q_q$ is the electric charge of the light
quark, $N_c = 3$, and $Q_t = 2/3$. For simplicity we have assumed in
Eq.~(\ref{eq:dq}) that the coupling of the Higgs to up and down quarks
is CP conserving. Note further that both $d_q$ and $\tilde d_q$ vanish
identically if the Higgs does not couple to the first generation of
quarks.

The two-loop matching correction of the Weinberg operator in
Eq.~(\ref{eq:LeffN}) has been calculated in~\cite{Dicus:1989va},
giving
\begin{equation} \label{eq:whigh}
w (\mu_W) = \frac{g_s}{4} \frac{\alpha_s}{(4 \pi)^3} \, \sqrt{2} G_F
\,  \kappa_t \tilde \kappa_t  \, f_3 (x_{t/h}) \,, 
\end{equation}
where\footnote{For $x \to \infty$, one finds that $f_3 (x) = 1 -
  {1}/{3x} +{\cal O} (1/x^2)$, while the measured values of $m_t$ and
  $M_h$ numerically lead to $f_3 (x_{t/h}) \simeq 0.87$. }
\begin{equation}
f_3 (x) = 4 x^2 \int_0^1 \! dv \int_0^1 \! du \; \frac{u^3 v^3 \left (1-v \right ) }
{\left [ x \hspace{0.5mm} v \left ( 1- u v \right ) + \left  (1- u \right ) \left (1 - 
v \right ) \right ]^2} \,. 
\end{equation}
Notice that the coefficient $w$ of the Weinberg operator depends only
on the top-quark couplings. The neutron EDM thus provides a constraint
on the product $\kappa_t\tilde \kappa_t$ even if the Higgs boson does
not couple to the first generation of fermions. This constraint is
complementary to the bounds from the Higgs production cross section at
the LHC, which is proportional to the sum of $\kappa_t^2$ and
$\tilde\kappa_t^2$ with appropriate weights (see
Sec.~\ref{sec:onshell:Higgs}).

The contributions of the EDM, CEDM, and Weinberg operators to the
neutron EDM are then given by~\cite{Pospelov:2005pr} (see also
\cite{Demir:2002gg,Kamenik:2011dk})
\begin{equation}
\begin{split}
\frac{d_n}{e} & = (1.0 \pm 0.5) \left \{ 1.4\left  [\frac{d_d (\mu_H)}{e} - 
0.25 \, \frac{d_u  (\mu_H)}{e} \right ] + 1.1 \left  [\tilde d_d (\mu_H) + 
0.5 \, \tilde d_u  (\mu_H) \right ] \right \} \\[3mm] 
& \phantom{xx}+  (22 \pm 10) \cdot 10^{-3} \, {\rm GeV} \, w (\mu_H) \,, 
\end{split}
\end{equation} 
where $\mu_H = 1 \, {\rm GeV}$ is a hadronic scale. The RG evolution
of the coefficients $d_q$, $\tilde d_q$, and $w$ from the weak to the
hadronic scale is given in App.~\ref{app:neutronEDM:RG}. After
performing the RG resummation we find the following numerical estimate
for the CP-violating Higgs-top coupling contribution to the neutron
EDM,
\begin{equation} \label{dnefinal} 
\begin{split}
\frac{d_n}{e} & = \Big \{ (1.0 \pm 0.5) \left [ - 5.3 \hspace{0.5mm}
  \kappa_q \tilde\kappa_t  + 5.1 \cdot 10^{-2}  \hspace{0.5mm}
  \kappa_t \tilde\kappa_t  \right ] \\[3mm] &  
\phantom{xxx} + (22  \pm 10 ) \, 1.8 \cdot 10^{-2}  \hspace{0.5mm}
\kappa_t  \tilde\kappa_t  \Big \} \cdot 10^{-25} \, {\rm cm} \,. 
\end{split}
\end{equation}
For simplicity we have identified here the modifications of the
CP-conserving up- and down-quark couplings, $\kappa_q = \kappa_u =
\kappa_d$. This shows that the contribution of the Weinberg operator
(which is proportional to the combination $\kappa_t \tilde\kappa_t $)
is numerically subdominant to the quark EDM and CEDM
contributions. Taking as an illustration the SM values for the
CP-conserving couplings, i.e.~$\kappa_t=\kappa_q=1$, the 95\% CL upper
bound on the neutron EDM~\cite{Baker:2006ts}
\begin{equation}  \label{eq:bestdn}
\left | \frac{d_n}{e} \right | < 2.9 \cdot 10^{-26} \, {\rm cm} \,,
\end{equation} 
leads  to
\begin{equation} \label{eq:kappatnEDM}
|\tilde\kappa_t|  < [0.03,0.10]\,,
\end{equation}
which is weaker by almost an order of magnitude than the constraint
(\ref{eq:kappateEDM}) arising from the electron EDM.

\subsection{EDM  of mercury}

The EDMs of diamagnetic atoms, i.e.~atoms where the total angular
momentum of the electrons is zero, also provide important tests of CP
violation of the Higgs-quark interactions. Presently, the most
stringent constraint in the diamagnetic sector comes from the limit on
the EDM of mercury (Hg). The dominant contribution to $d_{\rm Hg}$
arises from CP-odd pion nucleon interactions involving the isovector
channel ($g_{\pi N N}^{(1)}$), while isoscalar contributions ($g_{\pi
  N N}^{(0)}$) are accidentally small and effects related to the
Weinberg operator are chirally suppressed (see~\cite{Jung:2013hka} for
a comprehensive discussion of the theoretical errors plaguing the
prediction of $d_{\rm Hg}$). Including only effects associated with
the CP-odd pion nucleon coupling $g_{\pi N N}^{(1)}$, one
obtains~\cite{Pospelov:2005pr}
\begin{equation}
\frac{d_{\rm Hg}}{e} \simeq -1.8 \cdot 10^{-4} \left ( 4^{+8}_{-2}
\right ) \left (\tilde d_u (\mu_H) - \tilde d_d (\mu_H)  \right )\,. 
\end{equation}
Numerically, we find
\begin{equation} \label{dHgefinal} 
\frac{d_{\rm Hg}}{e} = - \left ( 4^{+8}_{-2} \right ) \, \Big [
  3.1 \hspace{0.5mm} \tilde\kappa_t  - 3.2\cdot
  10^{-2}  \hspace{0.5mm} \kappa_t  \tilde\kappa_t  \Big ] \cdot
10^{-29} \, {\rm cm} \,,  
\end{equation}
which should be compared to the 95\% CL bound~\cite{Griffith:2009zz}
\begin{equation}  \label{eq:bestdHg}
\left | \frac{d_{\rm Hg}}{e}  \right | < 3.1 \cdot 10^{-29} \, {\rm cm} \,,
\end{equation}
when deriving limits on $\kappa_t$ and $\tilde \kappa_t$. 

\section{Constraints from Higgs production and decay}
\label{sec:onshell:Higgs}

The CP-violating Higgs couplings affect the production cross sections
and decay branching ratios of the Higgs. One can devise targeted
search strategies optimized to the specifics of the kinematical
distributions induced by the CP-violating couplings
\cite{Berge:2008wi,Berge:2008dr,Berge:2011ij, Fermilab:CPV,Htt:CPV}.
Here we will be concerned only with the modifications of the total
rates, focusing primarily on the couplings of the Higgs to the top,
while the effect of bottom and tau couplings will be discussed in more
detail in Sec.~\ref{EDM:b:tau}.

Modifications of the Higgs-top couplings affect both the $gg \to h$ as
well as the $h \to \gamma \gamma$ vertex, which are generated at one
loop in the SM.  For the Higgs coupling to gluons one has the
following effective action
\begin{equation}\label{Lhgg}
V_{\rm eff} = - c_g
\,\frac{\alpha_s}{12\pi}\,\frac{h}{v}\,G_{\mu\nu}^a\,G^{\mu\nu,a} -
\tilde c_g \,\frac{\alpha_s}{8\pi}\,\frac{h}{v}\,
G_{\mu\nu}^a\,\widetilde G^{\mu\nu,a} \,. 
\end{equation}
At one loop the coefficients $c_g$ and $\tilde c_g$ are given by
\begin{equation}\label{matching}
c_g = \sum_{f=t,b} \kappa_f \,A(\tau_f) \,, \qquad \tilde c_g =
\sum_{f=t,b}  \tilde \kappa_f  \,B(\tau_f) \,, 
\end{equation}
where $\tau_f=4\hspace{0.25mm}m_{f}^2/M_h^2-i\varepsilon$ and 
\begin{equation}
A(\tau) = \frac{3\tau}{2}\,\Big[ 1 + (1-\tau)
  \arctan^2\frac{1}{\sqrt{\tau-1}} \Big] \,, \qquad B(\tau) = \tau
\arctan^2\frac{1}{\sqrt{\tau-1}} \,. 
\end{equation}
Since the top quark is sufficiently heavier than the Higgs boson,
$4\hspace{0.25mm}m_t^2\gg M_h^2$, it is a very good approximation to
use the asymptotic values $A(\infty)=B(\infty)=1$ in the case of a top
running in the loop. For light fermions, $\tau\ll1$, we have instead
{$A(\tau)\to -3 \tau/8 \hspace{0.5mm} \big [ \left (\ln \left (\tau/4
    \right )+i \pi \right )^2 -4 \big ]$ and $B(\tau)\to -
  \tau/4 \hspace{0.5mm} \left (\ln \left (\tau/4 \right )+i \pi \right
  )^2$ }.

The ratio of the cross sections for Higgs-boson production in
gluon-gluon fusion can now be written as
\begin{equation} \label{mug}
\mu_{gg} = \frac{\sigma (gg \to h)}{\sigma (gg \to h)_{\rm SM}} =
\big |\kappa_g \big |^2 +\big  |\tilde \kappa_g \big |^2\,, 
\end{equation}
with 
\begin{equation}  \label{kgluon}
\begin{split}
\kappa_g  &\equiv \frac{c_g}{c_{g,\rm SM}}=  \frac{\kappa_t  \,
  A(\tau_t) + \kappa_b A(\tau_b)}{\sum_{f=t,b} A(\tau_f)}\,,\\[2mm]  
\tilde \kappa_g   &\equiv  \frac{3}{2}\frac{\tilde c_g}{c_{g,\rm SM}}=
\frac{3}{2} \, \frac{\tilde \kappa_t  \,  B(\tau_t)+\tilde \kappa_b
  \,  B(\tau_b)}{\sum_{f=t,b} A(\tau_f)} \,.  
\end{split}
\end{equation}
Numerically, one has
\begin{equation} \label{eq:num:kappag} 
\kappa_g \simeq \left ( 1.05 -0.08 \hspace{0.5mm} i \right ) \kappa_t
- 0.05 + 0.08 \hspace{0.5mm} i \,, \qquad  
\tilde \kappa_g \simeq \left ( 1.60 -0.12 \hspace{0.5mm} i \right )
\tilde \kappa_t \,, 
\end{equation}
where we have set $\kappa_b =1$ and $\tilde \kappa_b = 0$ to obtain
the final expressions. The imaginary terms are the absorptive parts of
the amplitude that arise from virtual bottom quarks going
on-shell. This generates strong phases that do not flip sign under CP
conjugation. The only CP-violating contribution is therefore $\tilde
\kappa_g$, which is proportional to the fundamental CP-violating
coupling $\tilde \kappa_t$, as expected. Note that
\begin{equation} 
\mu_{gg}\simeq \kappa_t^2 +2.6\hspace{0.25mm} \tilde \kappa_t^2+
0.11\hspace{0.25mm}  \kappa_t \left (\kappa_t-1 \right )\,, 
\end{equation} 
so that the CP-violating Higgs-top coupling always enhances the signal
strength compared to the case of purely CP-conserving couplings.

Similarly, we can define the effective action for the Higgs coupling
to two photons
\begin{equation}\label{Lhgammagamma}
V_{\rm eff} = - c_\gamma
\,\frac{\alpha}{\pi}\,\frac{h}{v}\,F_{\mu\nu}\,F^{\mu\nu} - \tilde
c_\gamma \,\frac{3\alpha}{2\pi}\,\frac{h}{v}\, F_{\mu\nu}\,\widetilde
F^{\mu\nu} \,, 
\end{equation}
where
\begin{equation}\label{matching2}
c_\gamma = A_W + \sum_{f=t,b,\tau} \frac{N_c(f)}{6} \,
Q_f^2  \hspace{0.25mm}\kappa_f \,A(\tau_f) \,, \qquad  
\tilde c_\gamma =  \sum_{f=t,b,\tau} \frac{N_c(f)}{6} \,
Q_f^2  \hspace{0.25mm}\tilde \kappa_f  \,B(\tau_f) \,, 
\end{equation}
with 
\begin{equation}
A_W= - \frac{1}{8} \left [ 2 + 3 \hspace{0.25mm} \tau_W +
  3 \hspace{0.25mm} \tau_W  \left (2-\tau_W \right )
  \arctan^2\frac{1}{\sqrt{\tau_W-1}} \right ] \,.  
\end{equation}
and $\tau_W=4m_{W}^2/M_h^2-i\varepsilon$. Here $N_c(t) =N_c(b)=3$,
$N_c(\tau) =1$, and $\widetilde F^{\mu\nu} =
\frac12\epsilon^{\mu\nu\alpha\beta}\, F_{\alpha\beta}$ is the
electromagnetic dual field-strength tensor. The modification of the
signal strength for Higgs decays into two photons is parametrized by
\begin{equation}\label{eq:mugammagamma}
\mu_{\gamma\gamma}=\frac{\Gamma(h\to \gamma\gamma)}{\Gamma(h\to
  \gamma\gamma)_{\rm SM}}=\big |\kappa_\gamma \big |^2+ \big
|\tilde\kappa_\gamma \big |^2, 
\end{equation}
where
\begin{equation}  \label{kgamma}
\begin{split}
\kappa_\gamma & \equiv \frac{c_\gamma}{c_{\gamma, \rm SM}}=  \frac{A_W
  + \frac{2}{9}   \hspace{0.25mm}  \kappa_t  \hspace{0.25mm}
  A(\tau_t) + \frac{1}{18} \hspace{0.25mm} \kappa_b  \hspace{0.25mm}
  A(\tau_b) + \frac{1}{6}  \hspace{0.25mm}
  \kappa_\tau  \hspace{0.25mm}  A(\tau_\tau) }{A_W +
  \frac{2}{9} \hspace{0.25mm}  A(\tau_t) +
  \frac{1}{18} \hspace{0.25mm} A(\tau_b) + \frac{1}{6} \hspace{0.25mm}
  A(\tau_\tau) } \,, \\[2mm]  
\tilde \kappa_\gamma &  \equiv \frac{3}{2} \frac{\tilde
  c_\gamma}{c_{\gamma, \rm SM}} = \frac{\frac{1}{3} \hspace{0.25mm}
  \tilde \kappa_t   \hspace{0.25mm} B(\tau_t) +
  \frac{1}{12}  \hspace{0.25mm}  \tilde \kappa_b  \hspace{0.25mm}
  B(\tau_b) + \frac{1}{4}  \hspace{0.25mm}  \tilde
  \kappa_\tau \hspace{0.25mm}  B(\tau_\tau) }{A_W  +
  \frac{2}{9} \hspace{0.25mm}  A(\tau_t) +
  \frac{1}{18}  \hspace{0.25mm} A(\tau_b) +
  \frac{1}{6}  \hspace{0.25mm} A(\tau_\tau) } \,.  
\end{split}
\end{equation}
In the SM the $h\to \gamma\gamma$ decay width is dominated by $W$
bosons running in the loop, which gives $A_W \simeq -1.04$ using
$\tau_W \simeq1.63$. Assuming that the only modifications are in the
Higgs-top couplings (and thus setting $\kappa_b=\kappa_\tau=1$ and
$\tilde \kappa_b=\tilde \kappa_\tau=0$) one arrives at
\begin{equation} \label{eq:kgammagamma:num}
\kappa_\gamma \simeq  -0.28 \hspace{0.5mm} \kappa_t   + 1.28 \,,
\qquad \tilde \kappa_\gamma \simeq  -0.43 \hspace{0.5mm} \tilde
\kappa_t  \,. 
\end{equation}
Notice that the CP-violating coupling $\tilde \kappa_t$ always gives a
positive contribution to $\mu_{\gamma\gamma}$ compared to the
CP-conserving case. While the sign of $\kappa_t$ is not very important
for $\mu_{gg}$ as it only affects the numerically sub-leading
interference with the bottom-quark contribution, for
$\mu_{\gamma\gamma}$ the sign of $\kappa_t$ is crucial. Given the
destructive interference between the $W$-boson and the top-quark loop,
positive values of $\kappa_t$ diminish $\mu_{\gamma\gamma}$, while a
negative $\kappa_t$ has the opposite effect on $\mu_{\gamma\gamma}$.

The precise meaning of these modifications for different Higgs signal
strengths depends on the particular channel considered. For instance,
the inclusive Higgs di-photon rate is dominated by the gluon-gluon
fusion cross section, so that the modified signal strength due to
non-standard Higgs-top couplings is simply $\mu_{\gamma\gamma,\rm
  incl}=\mu_{gg} \hspace{0.25mm} \mu_{\gamma\gamma}$, with $\mu_{gg}$
given in Eqs.~\eqref{mug}, \eqref{eq:num:kappag} and
$\mu_{\gamma\gamma}$ in Eqs.~\eqref {eq:mugammagamma},
\eqref{eq:kgammagamma:num}. For the case where the bottom and tau
couplings are modified one must, however, take into account the
changes in the total rate. We will come back to this point in
Sec.~\ref{EDM:b:tau}.

\section{Combined constraints on top couplings}
\label{combined-top}

We next combine the EDM and Higgs signal-strength constraints on the
CP-violating Higgs-top coupling. We use the results of a global fit to
Higgs production channels performed by experimental collaborations,
where the effective $gg \to h$ and $h\to\gamma\gamma$ couplings are
left to vary freely. All the remaining couplings are set to their SM
values. This corresponds to our case, where only the couplings of the
top quark to the Higgs are modified. The ATLAS collaboration measures
$\kappa_g = 1.04 \pm 0.14$, $\kappa_\gamma = 1.20\pm 0.15$
\cite{Aad:2013wqa}, and the CMS collaboration obtains
$(\kappa_g,\kappa_\gamma) = ( 0.83,0.97)$ for the best-fit value,
while the 95\% CL regions for each of these couplings separately are
$\kappa_g\in [0.63,1.05]$ and $\kappa_\gamma\in [0.59,1.30]$
\cite{CMS-PAS-HIG-13-005}. A naive weighted average then gives
\begin{equation}
\kappa_{g,\rm WA}=0.91\pm0.08\,,\qquad \kappa_{\gamma,\rm WA}=
1.10\pm0.11\,,
\end{equation} 
for the experimental world averages. In the experimental analyses
CP-conserving couplings to the Higgs are assumed. With the addition of
CP-violating couplings the efficiencies for different Higgs production
and decay channels can change in principle. For the moment, we ignore
this subtlety and simply set $\kappa_{g,\rm WA}^2 = |\kappa_g|^2 +
|\tilde\kappa_g|^2$ and $\kappa_{\gamma,\rm WA}^2 = |\kappa_\gamma|^2
+ |\tilde\kappa_\gamma|^2$ in our numerical estimates of the
experimental constraints. This approximate treatment can easily be
improved once more information on the dependence of the efficiencies
on the assumption of CP conservation is available from experiments. We
also neglect the correlations between the measurements of $\kappa_g$
and $\kappa_\gamma$, which is a good approximation
\cite{Aad:2013wqa,CMS-PAS-HIG-13-005}.

\begin{figure}[!t]
\begin{center}
\vspace{-5mm}
\includegraphics[height=0.42 \textwidth]{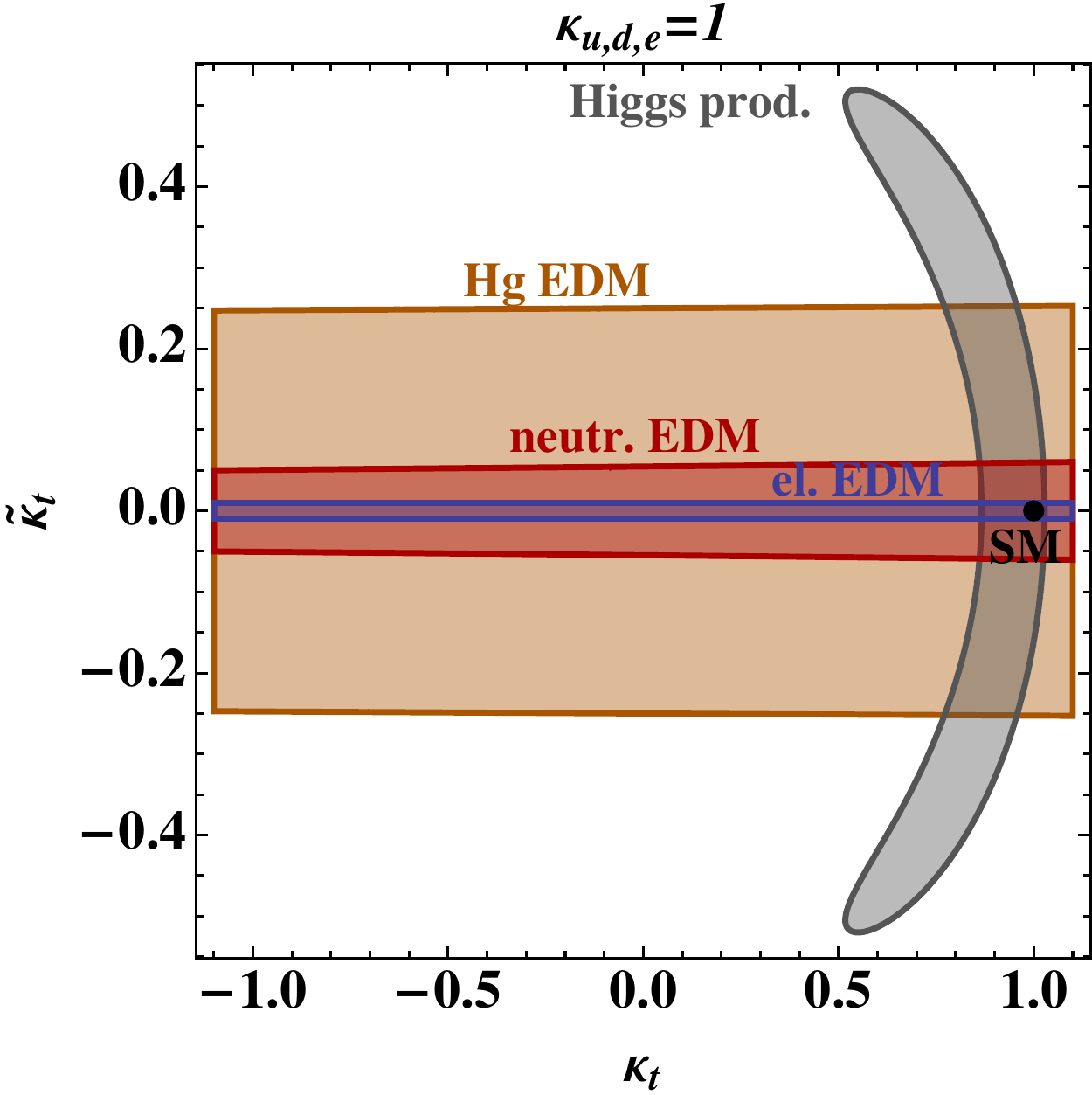} \qquad 
\includegraphics[height=0.42 \textwidth]{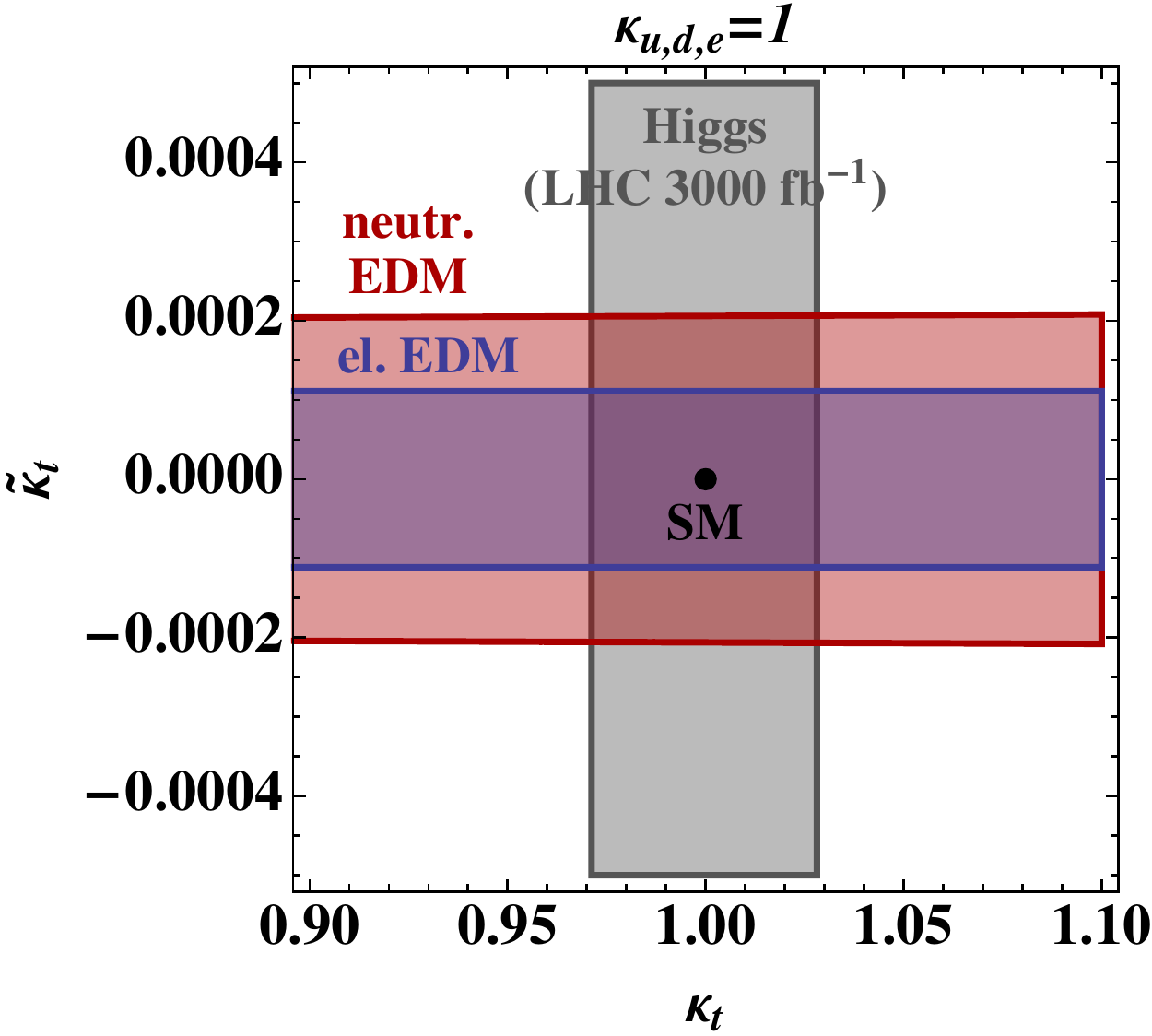}
\caption{\label{fig2} Left: Present constraints on $\kappa_t $ and
  $\tilde\kappa_t $ from the electron EDM (blue), the neutron
  EDM~(red), the mercury EDM (brown), and Higgs physics (gray). Right:
  Projected future constraints on $\kappa_t $ and $\tilde\kappa_t $,
  see text for details. }
\end{center}
\end{figure}

The present constraints on $\kappa_t $ and $\tilde\kappa_t $ are shown
in Fig.~\ref{fig2} (left). The regions allowed by the electron EDM,
neutron EDM, mercury EDM, and collider constraints are colored in
blue, red, brown, and gray, respectively, while the black point
corresponds to the SM prediction. The constraints resulting from the
EDM of the neutron and mercury employ the central values of the matrix
elements in Eqs.~(\ref{dnefinal}) and~(\ref{dHgefinal}).  Note that
the corrections to the $gg \to h$ and $h\to \gamma\gamma$ vertices
scale differently with $\kappa_t$ and $\tilde \kappa_t$ and thus
provide complementary constraints. The Higgs measurements are precise
enough that they already by themselves constrain the CP-violating
modification of the Higgs-top coupling to be below $\tilde
\kappa_t\lesssim {\mathcal O}(0.5)$. The EDM constraints shrink the
allowed region further to $\tilde \kappa_t\lesssim {\mathcal
  O}(0.01)$.

The right panel in Fig.~\ref{fig2} shows the prospects of the
constraints. In order to obtain the plot we have assumed that $|d_e/e|
< 10^{-30} \, {\rm cm}$ \cite{Hewett:2012ns}, a factor of $90$
improvement over the current best limit (\ref{eq:bestde}), and that
$|d_n/e| < 10^{-28} \, {\rm cm}$ \cite{Hewett:2012ns}, a factor of 300
improvement with respect to the present bound (\ref{eq:bestdn}).  Our
forecast for the future sensitivity of the Higgs production
constraints is based on the results of the CMS study with a projection
of errors to $3000$~fb${}^{-1}$, which assumed $1/\sqrt{{\cal L}}$
scaling of the experimental uncertainties with luminosity~${\cal L}$,
and also anticipates that the theory errors will be halved by then
\cite{Olsen:talk:Seattle}. In Fig.~\ref{fig2} we therefore take
$\kappa_g=1.00\pm0.03$ and {$\kappa_\gamma=1.00\pm0.02$} as the
possible future fit inputs (centered around the SM predictions).

Since the EDMs depend linearly on $\tilde \kappa_t$, the projected
order-of-magnitude improvements of the EDM constraints directly
translate to order-of-magnitude improvements of the bounds on $\tilde
\kappa_t$.  For instance, the electron EDM is projected to be
sensitive to values of $\tilde\kappa_t = {\cal O} (10^{-4})$ which
implies that one can probe scales up to $\Lambda = {\cal O} (25 \, {\rm
  TeV})$ for models (such as theories with top compositeness) where
{$\tilde\kappa_t \sim v^2/\Lambda^2$}.

\begin{figure}[!t]
\begin{center}
\vspace{-5mm}
\includegraphics[height=0.42 \textwidth]{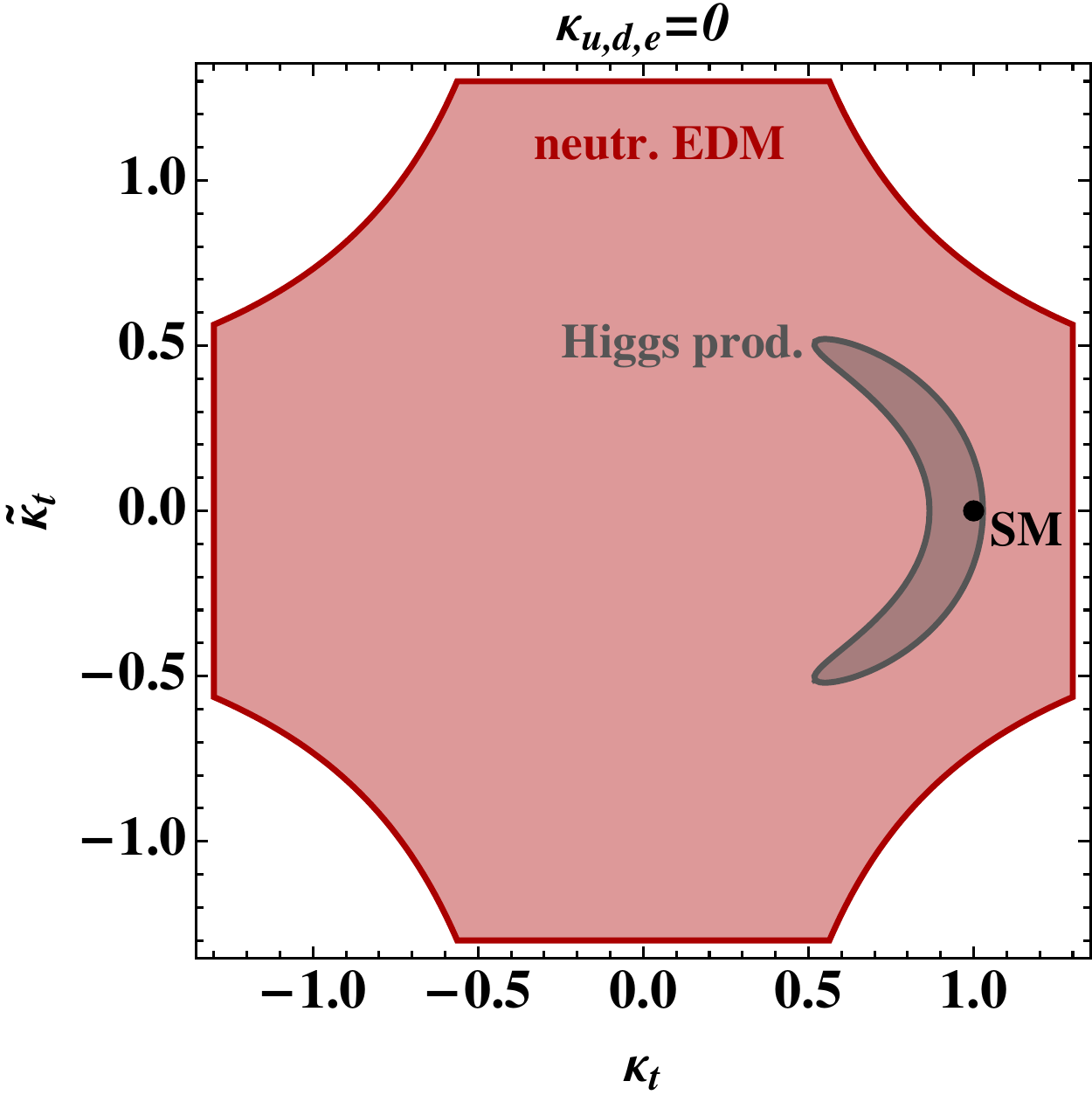} \qquad 
\includegraphics[height=0.42 \textwidth]{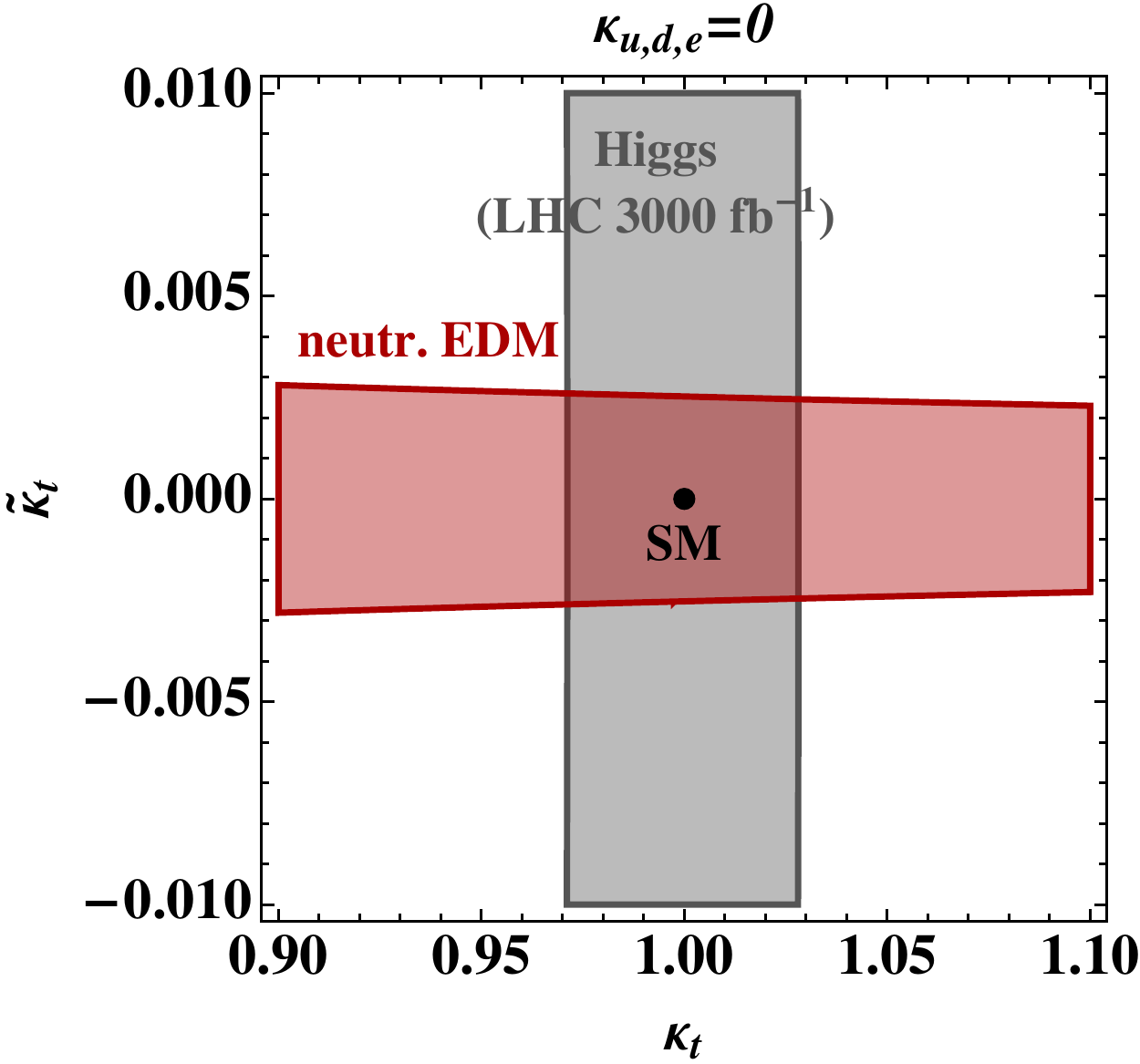}
\caption{\label{fig:top:3rdgen} Left: Present constraints on
  $\kappa_t$ and $\tilde\kappa_t $ from the neutron EDM~(red) and
  Higgs physics~(gray), assuming that the Higgs only couples to the
  third generation. Right: Projected future constraints on $\kappa_t $
  and $\tilde\kappa_t $, see text for details.  }
\end{center}
\end{figure}

Note that the above EDM constraints rely heavily on the assumption
that the Higgs couples to electrons, up, and down quarks. For
illustration we assumed that these couplings are the same as in the
SM. The possibility that the Higgs only couples to the
third-generation fermions cannot be ruled out from current Higgs
data. In this case there is no constraint from the electron EDM
which is proportional to $\kappa_e \hspace{0.125mm} \tilde
\kappa_t$. The neutron and mercury EDM are similarly dominated by
the quark EDMs and CEDMs which scale as
$\kappa_{u,d} \hspace{0.25mm} \tilde \kappa_t$. However, setting
$\kappa_{u,d}=0$ the constraints due to $d_n$ and $d_{\rm Hg}$ do
not vanish, because there is also a small contribution from the
Weinberg operator which scales as $\kappa_t \hspace{0.125mm} \tilde
\kappa_t$. In Fig.~\ref{fig:top:3rdgen} we show the constraints for
the limiting case where the Higgs only couples to the
third-generation fermions. We see that at present ${\mathcal O(1)}$
values of $\tilde \kappa_t$ are allowed by the constraint from the
neutron EDM. Assuming that only the Higgs-top
couplings are modified, the Higgs data are then more constraining than the neutron
EDM. This situation might change dramatically in the future with the
expected advances in the measurement of the neutron EDM. As
illustrated in Fig.~\ref{fig:top:3rdgen} (right), a factor 300
improvement in the measurement of $d_n$ will lead to ${\mathcal
  O}(10^{-3})$ constraints on $\tilde \kappa_t$, making the neutron
EDM as (or even more) powerful than the projected precision Higgs
measurements at a high-luminosity upgrade of the LHC.

\section{Constraints on bottom and tau couplings}
\label{EDM:b:tau}

In the following we analyze indirect and direct bounds on the
couplings between the Higgs and the other two relevant
third-generation fermions, i.e.~the bottom quark and the tau
lepton. In this case, the EDM constraints are suppressed by the small
bottom and tau Yukawa couplings, which renders the present indirect
limits weak.  However, given the projected order-of-magnitude
improvements in the experimental determinations of EDMs, relevant
bounds are expected to arise in the future. We will see that these
limits are complementary to the constraints that can be obtained via
precision studies of Higgs properties at a high-luminosity LHC.

\subsection{EDM constraints}
\label{sec:lightEDMs}

The bottom-quark and tau-lepton loop contributions to the electron EDM
are found from Eq.~(\ref{first}) after a simple replacement of charges
and couplings. The calculation of the hadronic EDMs, on the other
hand, is complicated by the appearance of large logarithms of the
ratios $x_{f/h} \equiv m_f^2/M_h^2$ with $f = b,\tau$. The structure
of the logarithmic corrections can be understood by evaluating
Eqs.~(\ref{eq:dq}) and (\ref{eq:whigh}) in the limit $x_{f/h} \to
0$. In the bottom-quark case, we find
\begin{equation} \label{eq:dsw}
\begin{split}
d_q (\mu_W) &\simeq -4 \hspace{0.25mm}  e  \hspace{0.25mm}
Q_q \hspace{0.25mm} N_c \hspace{0.25mm} Q_b^2 \,
\frac{\alpha}{(4\pi)^3} \hspace{0.25mm} \sqrt{2} G_F  \hspace{0.25mm}
m_q \, \kappa_q  \tilde\kappa_b   \,  x_{b/h} \left ( \ln^2 x_{b/h} +
\frac{\pi^2}{3} \right ) \,,\\[2mm] 
\tilde d_q (\mu_W) & \simeq -2  \,
\frac{\alpha_s}{(4\pi)^3} \hspace{0.25mm} \sqrt{2}
G_F  \hspace{0.25mm}  m_q \, \kappa_q  \tilde\kappa_b  \,  x_{b/h}
\left ( \ln^2 x_{b/h} + \frac{\pi^2}{3} \right )   \,, \\[2mm] 
w (\mu_W) & \simeq -g_s \hspace{0.25mm} \frac{\alpha_s}{(4 \pi)^3} \,
\sqrt{2} G_F \,  \kappa_b \tilde \kappa_b  \, x_{b/h} \left ( \ln
x_{b/h} + \frac{3}{2} \right) \,. 
\end{split}
\end{equation}
Here we have employed the asymptotic expansions $f_1 (x)= x \left (
\ln^2 x + \pi^2/3 \right) + {\cal O} (x^2)$ and $f_3 (x)= -4x \left (
\ln x + 3/2 \right) + {\cal O} (x^2)$ valid for $x \ll 1$. Note that
$\tilde d_q$ is proportional to $\alpha_s \ln^2 x_{b/h}$, whereas $w$
involves a term $\alpha_s \ln x_{b/h}$. This implies that only the
coefficient $\tilde d_q$ leads to a leading logarithmic (LL) effect,
while $w$ represents next-to-leading logarithmic (NLL) QCD
corrections. In order to obtain reliable results for $d_n$ and $d_{\rm
  Hg}$ the logarithmic QCD effects in Eq.~(\ref{eq:dsw}) have to be
resummed to all orders in the strong coupling constant using the full
machinery of RG-improved perturbation theory. We give details on this
RG calculation in App.~\ref{app:lightEDM:RG}. On the other hand, the
double logarithm in $d_q$ arises from QED corrections and thus
  does not need to be resummed. In the appendix we calculate the LL
QCD corrections to the quark EDM and show that they are larger than
the QED effects given above. Therefore we will include  both the leading 
QED and QCD contributions to~$d_q$ in our numerical analysis.

\begin{figure}[!t]
\begin{center}
\vspace{-5mm}
\includegraphics[height=0.42 \textwidth]{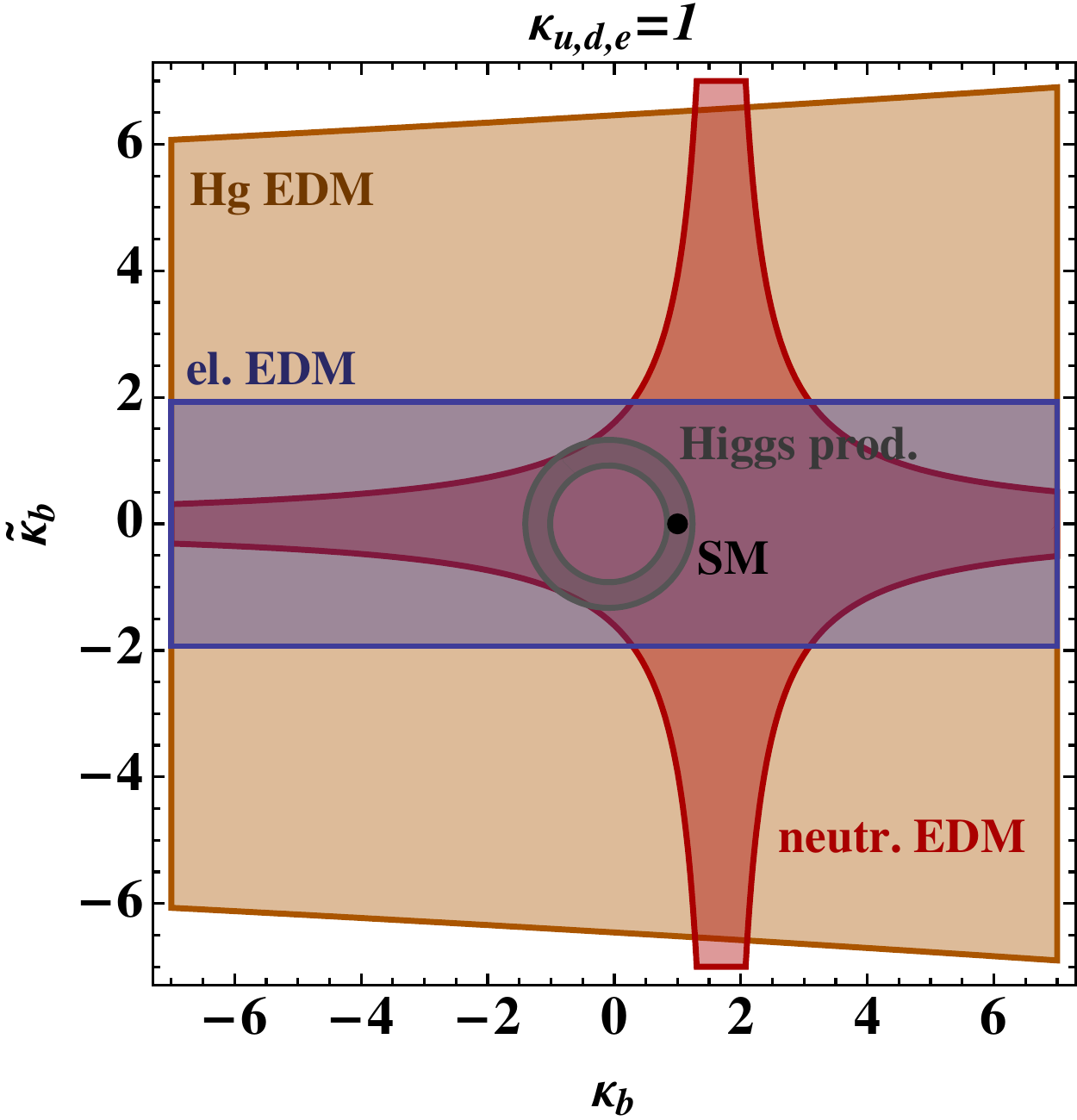} \qquad 
\includegraphics[height=0.42 \textwidth]{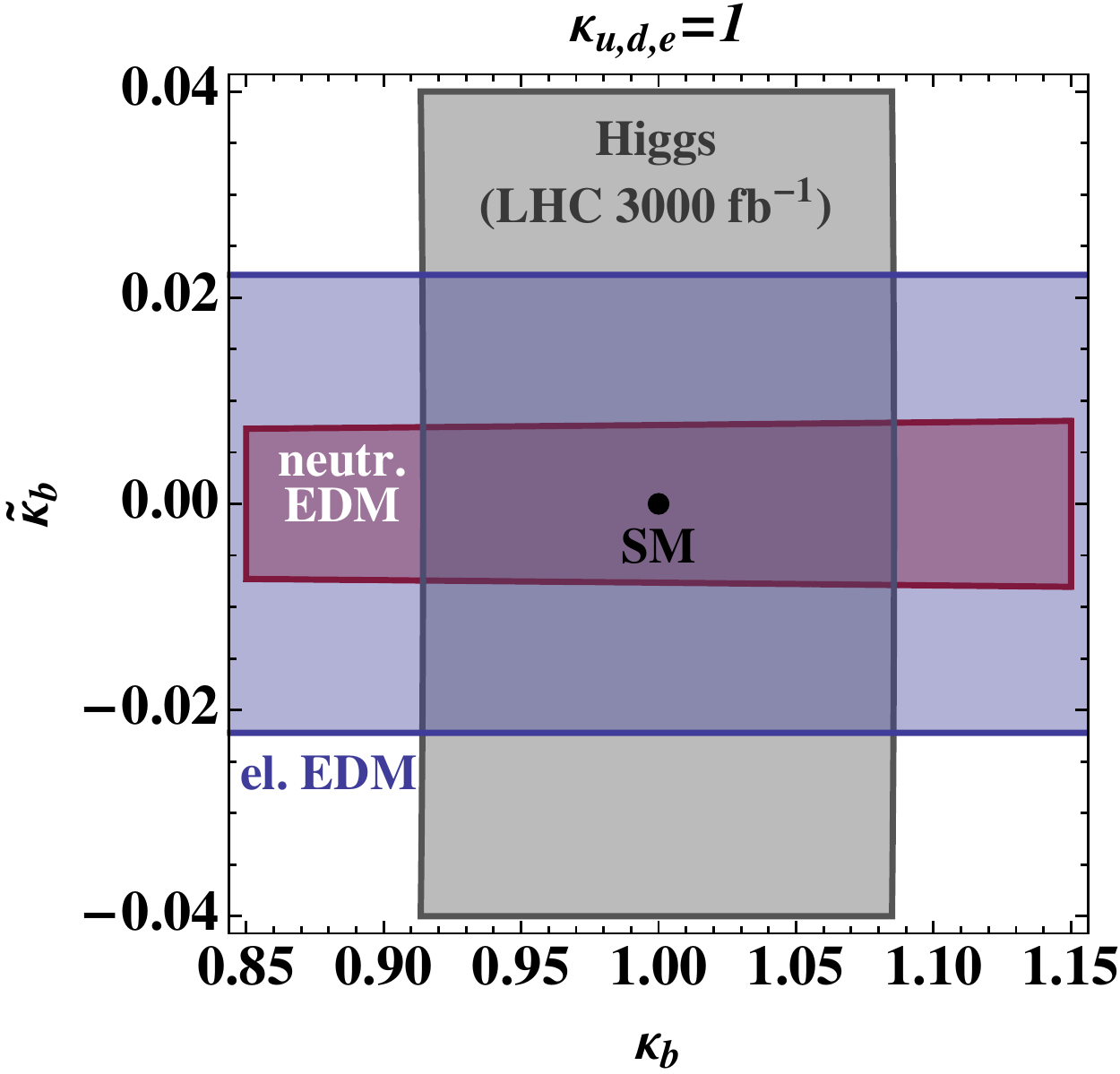}
\caption{\label{fig4} Left: Present constraints on $\kappa_b$ and
  $\tilde \kappa_b$ from the electron EDM (blue), the neutron
  EDM~(red), the mercury EDM (brown), and Higgs physics~(gray),
  restricting all other Higgs couplings to their SM values. Right: 
  Possible future constraints on $\kappa_b$ and $\tilde \kappa_b$, 
  see text for details. }
\end{center}
\end{figure}

Solving the relevant RG equations, we obtain the following approximate
expressions for the case of the Higgs-bottom couplings $\kappa_b$ and
$\tilde \kappa_b$: 
\begin{equation} 
\begin{split}
\frac{d_e}{e}  & =  4.5 \cdot 10^{-29} \, \tilde \kappa_b \, {\rm cm} \,, \\[2mm]
 \frac{d_n}{e} & = \Big \{ (1.0 \pm 0.5) \left [ -18.1 \hspace{0.5mm}
   \tilde \kappa_b + 0.15 \hspace{0.5mm} \kappa_b \tilde
   \kappa_b \right ] 
 + (22 \pm 10 ) \, 0.48
 \hspace{0.5mm} \kappa_b  \tilde\kappa_b  \Big \} \cdot
 10^{-27} \, {\rm cm} \,, \\[2mm] 
\frac{d_{\rm Hg}}{e} & = - \left (4^{+8}_{-2} \right ) \Big [0.12
  \hspace{0.5mm} \tilde \kappa_b - 1.1 \cdot
  10^{-3}  \hspace{0.5mm} \kappa_b \tilde \kappa_b \Big ] \cdot
10^{-29} \, {\rm cm} \,. 
\end{split}
\end{equation}
For the bottom-quark contribution to the electron EDM we take into
account logarithmic effects associated to the running of the
electromagnetic coupling constant by employing $\alpha \simeq 1/137$,
renormalized at zero-momentum transfer, which is appropriate for real
photon emission. In consequence, the above formula for $d_e$ is
obtained from the result for $d_q$ in Eq.~\eqref{eq:dsw} by replacing
$q\to e$ everywhere. Electromagnetic corrections describing operator
mixing are, on the other hand, not included.

In the case of modified Higgs-tau couplings $\kappa_\tau$ and $\tilde
\kappa_\tau$, we find 
\begin{equation} 
\begin{split}
\frac{d_e}{e} & =  3.7 \cdot 10^{-29} \, {\rm cm} \, \tilde
\kappa_\tau \,, \\[2mm]
 \frac{d_n}{e} & = \left (1.0 \pm 0.5 \right )
 22.3 \hspace{0.5mm} \tilde \kappa_\tau \cdot 10^{-29} \,
 {\rm cm} \,.
\end{split}
\end{equation}
Again no RG resummation of QED effects beyond the renormalization of
the electric charge has been performed here. Numerically, this
resummation is a ${\mathcal O}(10\%)$ correction for the neutron EDM,
which is clearly a sub-leading effect given the present hadronic
uncertainties in $d_n$. The expression for $d_n$ is obtained from
$d_q$, as given in Eq.~\eqref{eq:dsw}, by replacing $b\to \tau$ in all
subscripts and setting $N_c\to 1$, while in the case of $d_e$ one in
addition replaces $q\to e$ in all the subscripts. Since the mercury
EDM does not provide a meaningful constraint on the couplings
$\kappa_\tau$ and $\tilde \kappa_\tau$, we do not give an expression
for $d_{\rm Hg}/e$.

\subsection{Direct Higgs constraints}

The modified Higgs-bottom couplings induce corrections to the
effective $g g \to h$ and $h \to \gamma \gamma$ vertices of the
following form
\begin{equation}  \label{eq:neglect}
\begin{split}
 \kappa_g & \simeq \left (- 0.05 + 0.08 \hspace{0.5mm} i \right )
 \kappa_b + 1.05 -0.08 \hspace{0.5mm} i  \,, \qquad \hspace{5mm}
 \tilde \kappa_g \simeq \left ( -0.06  + 0.08 \hspace{0.5mm} i \right
 ) \tilde \kappa_b \,, \\[1mm] 
 \kappa_\gamma & \simeq \left ( 0.004 - 0.005 \hspace{0.5mm} i \right
 ) \kappa_b + 0.996 + 0.005 \hspace{0.5mm} i  \,, \qquad \tilde
 \kappa_\gamma \simeq \left ( 0.004  - 0.005 \hspace{0.5mm} i \right )
 \tilde \kappa_b \,. 
\end{split}
\end{equation}
The SM vertices in the two cases are dominated by the top-quark and
$W$-boson couplings to the Higgs, both of which are ${\mathcal O}(1)$,
and thus much larger than the SM bottom-quark Yukawa coupling $y_b =
{\cal O}( 0.02)$. As a result the corrections in $\kappa_{g,\gamma}$
and $\tilde \kappa_{g,\gamma}$ due to $\kappa_b \neq 1$ and $\tilde
\kappa_b \neq 0$ are sub-leading and can be neglected in our analysis
(we have checked this explicitly).

\begin{figure}[!t]
\begin{center}
\vspace{-5mm}
\includegraphics[height=0.42 \textwidth]{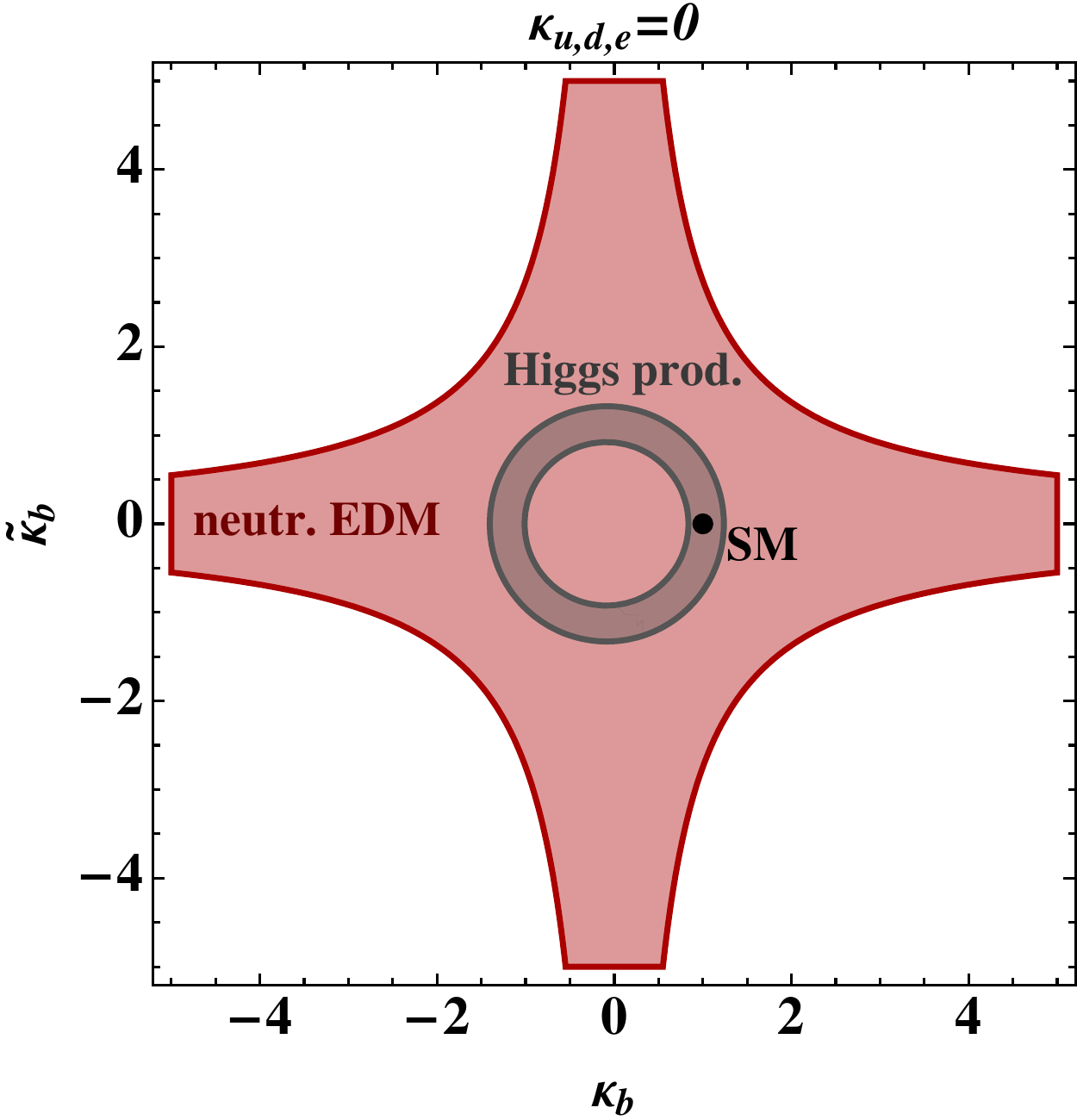} \qquad 
\includegraphics[height=0.42 \textwidth]{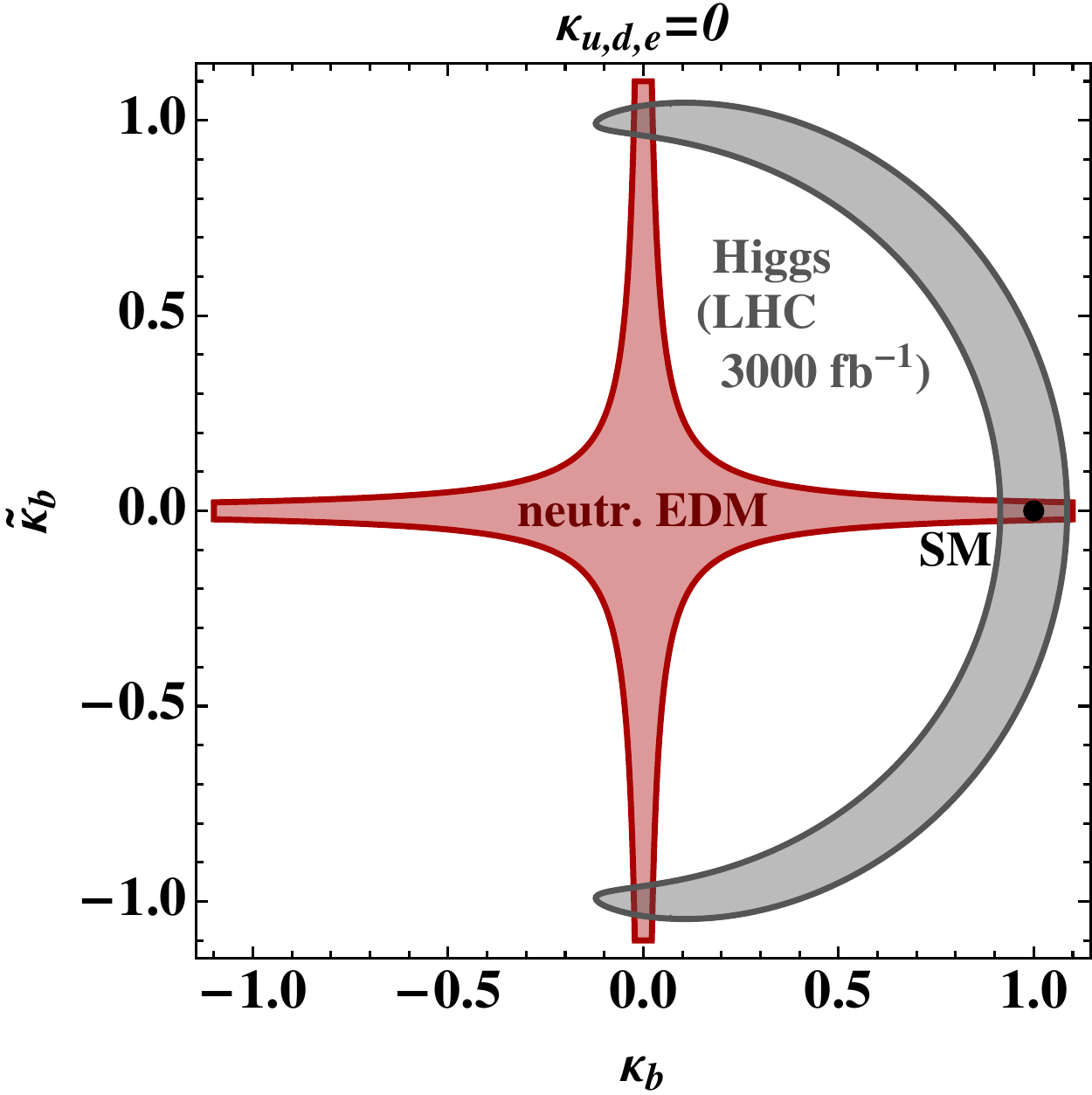}
\caption{\label{fig4:3rdgen} Left: Present constraints on $\kappa_b$
  and $\tilde \kappa_b$ from the neutron EDM~(red) and Higgs physics~(gray),
  assuming that the Higgs only couples to the third-generation
  fermions, $W$, and $Z$ bosons. Right: Possible future constraints on
  $\kappa_b$ and $\tilde \kappa_b$, see text for details.}
\end{center}
\end{figure}

The most significant change in the Higgs signals arises therefore from
the change in the total Higgs-decay width. Assuming that only the
Higgs coupling to bottom quarks is modified, the new total decay width
of the Higgs is
\begin{equation}
\Gamma=\left [1+\left (\kappa_b^2+\tilde \kappa_b^2-1 \right ){\rm
    Br}(h\to b\bar b)_{\rm SM}\right] \Gamma_{\rm SM}\, \,. 
\end{equation}  
This means that the $h\to b\bar b$ branching ratio is now
\begin{equation}\label{eq:Br:bb}
{\rm Br}(h\to b\bar b)=\frac{\left (\kappa_b^2+\tilde \kappa_b^2
  \right ){\rm Br}(h\to b\bar b)_{\rm SM}}{1+ \left (\kappa_b^2+\tilde
  \kappa_b^2-1 \right ){\rm Br}(h\to b\bar b)_{\rm SM}} \,, 
\end{equation}
while all the other Higgs-decay modes get rescaled to
\begin{equation}\label{eq:Br:X}
{\rm Br}(h\to X)=\frac{{\rm Br}(h\to X)_{\rm SM}}{1+\left
  (\kappa_b^2+\tilde \kappa_b^2-1 \right ){\rm Br}(h\to b\bar b)_{\rm
    SM}} \,, 
\end{equation}
where $X\ne b\bar b$. As inputs we use the naive averages of the ATLAS
\cite{ATLAS-CONF-2013-034} and CMS collaborations
\cite{CMS-PAS-HIG-13-005} in different Higgs-decay channels
\begin{equation}
\begin{split}
\hat \mu_{b\bar b}&=0.72\pm0.53\,, \quad \hat\mu_{\tau\bar
  \tau}=1.02\pm0.35\,,
\quad\hat\mu_{\gamma\gamma}=1.14\pm0.20\,,\\[2mm] 
&\qquad\hat \mu_{WW}=0.78\pm0.17\,, \quad \hat \mu_{ZZ}=1.11\pm0.23\,,
\end{split}
\end{equation} 
where $\hat\mu_X\equiv [\sigma(pp\to h)\hspace{0.5mm}{\rm Br}(h\to
  X)]/[\sigma(pp\to h)\hspace{0.5mm}{\rm Br}(h\to X)]_{\rm SM}$
denotes the signal strengths. We work in the limit where the
Higgs couplings to the $W$ and $Z$ bosons are the SM ones. 
We keep the effect of $\kappa_b, \tilde \kappa_b$ in the $g g \to h$ 
and $h\to \gamma\gamma$ vertices~$\big($cf.~Eq.~(\ref{eq:neglect})$\big)$, 
where the former interaction also modifies the Higgs production cross section. 
Up to these sub-leading corrections the changes in the signal strengths are 
the same as in the corresponding branching ratios, Eqs.~\eqref{eq:Br:bb},
\eqref{eq:Br:X}, with $\hat\mu_X={\rm Br}(h\to X)/{\rm Br}(h\to X)_{\rm SM}$.

The resulting direct constraints in the $\kappa_b$--$\tilde \kappa_b$
plane are displayed in Fig.~\ref{fig4}. As shown in the left panel,
the present restrictions from Higgs physics carve out a ring-like
allowed region that corresponds to effects of ${\cal O} (1)$ in
$\kappa_b$ and $\tilde \kappa_b$. We also see that the EDMs currently
impose even weaker bounds, with the strongest limit coming from
$d_n$. However, the relative strength of the two sets of constraints
is expected to change in the future, as illustrated in the right
panel. For our forecast we use the CMS projections for a $h\to b\bar
b$ coupling measurement with 3000 fb${}^{-1}$ of integrated
luminosity~\cite{Olsen:talk:Seattle}, assuming that this bounds the
combination $\kappa_b^2+\tilde \kappa_b^2$. Including the constraints
from the projected measurements of the $gg \to h$ and $h\to
\gamma\gamma$ vertices breaks the symmetry between $\kappa_b$ and
$\tilde \kappa_b$, so that only part of the ring-like region survives
(we used the SM values for the central values of the hypothetical
measurements). This limits the size of possible modifications in
$\kappa_b$ to ${\cal O} (0.05)$. Complementary information is obtained
in such a future scenario from the envisioned high-precision
measurements of the electron and neutron EDM, which might allow to
probe values of the CP-violating coefficient $\tilde \kappa_b$ down to
${\cal O} (10^{-2})$.

While the EDM constraints depicted in Fig.~\ref{fig4} assume that the
Higgs couples to first-generation fermions with SM strength,
meaningful EDM constraints on $\tilde \kappa_b$ can even emerge if
$\kappa_{u,d}=0$. In fact, as illustrated in Fig.~\ref{fig4:3rdgen},
the neutron EDM probes $\tilde \kappa_b$ through the Weinberg operator
also if the Higgs couples only to the third generation. While at
present~(left panel) no relevant constraint can be derived in such a
case, extracting a limit on $\tilde\kappa_b$ of ${\mathcal O}(0.1)$
may be possible in the future (right panel) if $\kappa_b$ is SM-like.
This feature again highlights the power of low-energy EDM measurements
in probing new sources of CP violation.

\begin{figure}[!t]
\begin{center}
\vspace{-5mm}
\includegraphics[height=0.42 \textwidth]{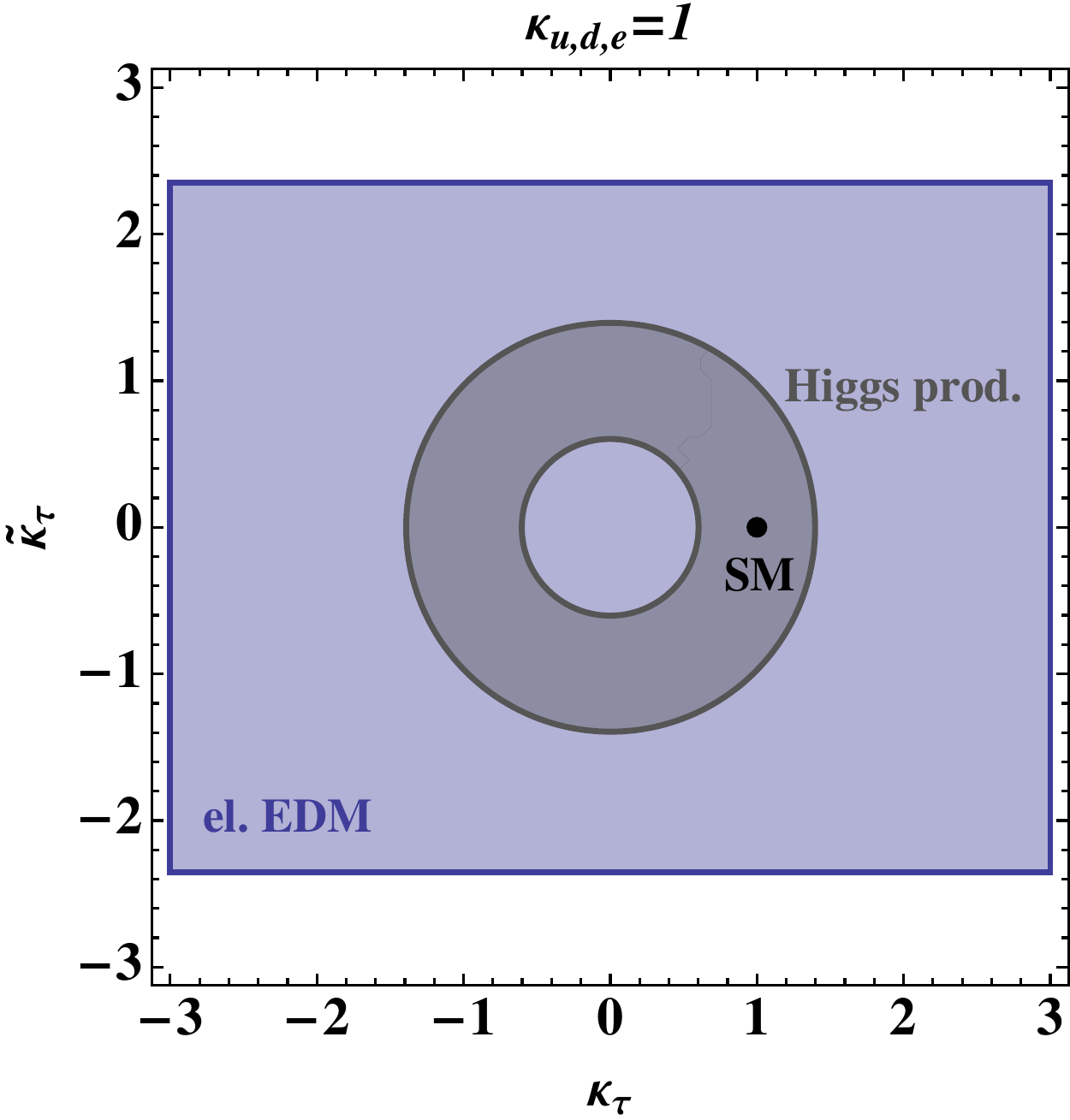} \qquad 
\includegraphics[height=0.42 \textwidth]{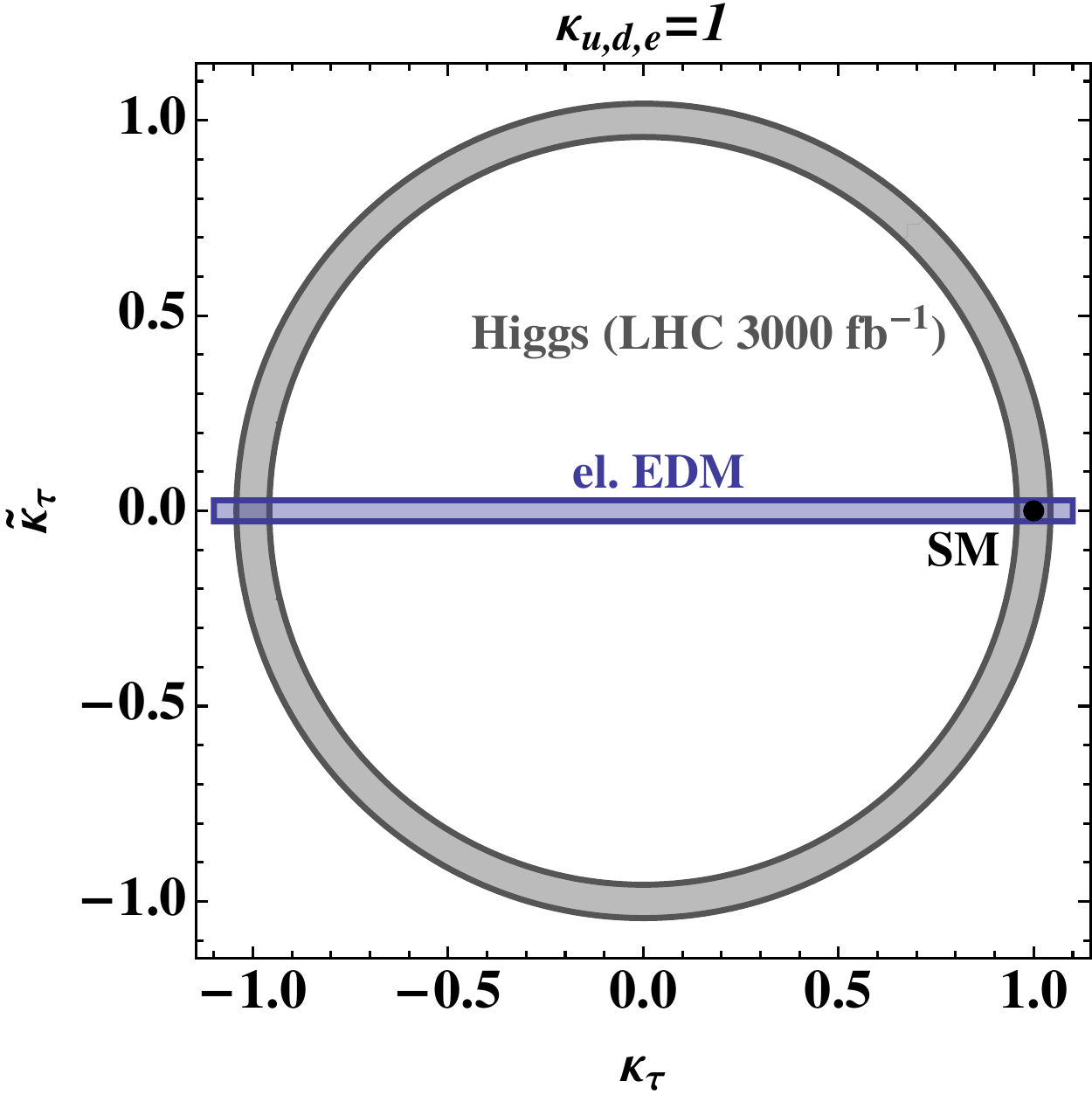}
\caption{\label{fig5} Left: Present constraints on $\kappa_\tau$ and
  $\tilde \kappa_\tau$ from the electron EDM (blue) and Higgs
  production~(gray), assuming SM values for the remaining Higgs
  couplings. Right: Possible future constraints on $\kappa_\tau$ and
  $\tilde \kappa_\tau$, see text for details. }
\end{center}
\end{figure}

Modifying the Higgs-tau couplings changes the effective $h\to
\gamma\gamma$ vertex. The induced shifts are parametrized by
\begin{equation} 
 \kappa_\gamma  \simeq \left ( 0.004 - 0.003 \hspace{0.5mm} i \right )
 \kappa_\tau + 0.996 + 0.003 \hspace{0.5mm} i  \,, \qquad 
\tilde \kappa_\gamma \simeq \left ( 0.004  - 0.003 \hspace{0.5mm} i
\right ) \tilde \kappa_\tau \,. 
\end{equation}
Similar to the case of Higgs couplings to bottom quarks, the
corrections to $\kappa_\gamma$ and $\tilde \kappa_\gamma$ are
suppressed by the small tau Yukawa coupling, $y_\tau = {\cal O}
(0.01)$. The main effect is therefore the rescaling of the total decay
widths, as in Eqs.~\eqref{eq:Br:bb},~\eqref{eq:Br:X}, but replacing
$b\to \tau$. The resulting constraints in the $\kappa_\tau$--$\tilde
\kappa_\tau$ plane are displayed in Fig.~\ref{fig5}, with the left
panel showing the current bounds, and the right panel the
extrapolation to $3000$~fb${}^{-1}$ of integrated luminosity, using
again~\cite{Olsen:talk:Seattle}. One observes that even the projected
precision of $2\%$ on $\kappa_\gamma$ will not suffice to break the
symmetry between $\kappa_\tau$ and $\tilde \kappa_\tau$ and the
ring-like bound persists, allowing for potentially ${\mathcal O}(1)$
values of the CP-violating modification $\tilde \kappa_\tau$. While at
present the EDMs lead to a bound $|\tilde\kappa_\tau| \lesssim 2$, of
the same order but slightly weaker than the collider constraint,
assuming a factor of $90$ improvement in the determination of the
electron EDM will change the situation, as it will make values $\tilde
\kappa_\tau = {\cal O} (10^{-2})$ accessible. Direct searches at the
LHC using angular correlations in the $h\to\tau\bar \tau$ channel may
be capable to probe $\tilde \kappa_\tau$ values of ${\cal O}
(0.1)$~\cite{Berge:2008wi, Berge:2008dr,Berge:2011ij, Fermilab:CPV},
and are thus less powerful than the indirect bounds. Unlike the
constraint from the electron EDM, direct bounds, however, do not
depend on the assumption $\kappa_e =1$.

\section{Conclusions}
\label{conclusions}

The LHC discovery of the Higgs boson furnishes new opportunities in
the search for physics beyond the SM. Since in the SM the Higgs
couplings to both gauge bosons and fermions are uniquely fixed in
terms of the corresponding masses, finding a significant deviation
from this simple pattern would constitute a clear signal of NP. In
fact, a major experimental effort is directed towards determining the
structure of the Higgs sector including its CP properties by measuring
the various decay rates of the new boson as accurately as
possible. While the current LHC results favor purely scalar-like
Higgs-gauge boson interactions, searches for CP violation in fermionic
Higgs decays are still in their fledgling stages.

In this article we have emphasized the complementarity between
high-$p_T$ and low-energy precision measurements in extracting
information about the CP properties of the Higgs-boson couplings to
third-generation fermions. In the case of the Higgs-top couplings we
find that the existing data on Higgs production and decay are already
precise enough to constrain the CP-violating modification to $\tilde
\kappa_t \lesssim {\cal O} (0.5)$. The present constraints arising
from the EDMs shrink the allowed region further to $\tilde \kappa_t =
{\cal O} (0.01)$, if SM couplings of the Higgs to the first generation
fermions are assumed. At a high-luminosity LHC and the next generation
of EDM experiments it should be possible to improve the above limits
on CP violation in the Higgs-top coupling significantly. Our analysis
shows that while at the 14~TeV LHC with $3000 \, {\rm fb}^{-1}$ of
integrated luminosity a sensitivity of $\tilde \kappa_t = {\cal O}
(10^{-2})$ can be reached, the electron EDM is projected to be
sensitive to values down to $\tilde \kappa_t = {\cal O}
(10^{-4})$. Such a precision will allow to indirectly probe for NP
scales up to $\Lambda = {\cal O} (25 \, {\rm TeV})$ in models, such as
theories with top compositeness, that predict $\tilde \kappa_t \sim
v^2/\Lambda^2$.

The above EDM bounds on $\tilde \kappa_t$ only apply under the
assumption that the electron and the down- and up-quark Yukawa
couplings take their SM values.  This requirement can be avoided, however. 
The constraints due to the neutron and mercury EDM  do not vanish even if
the Higgs boson couples only to the third generation of fermions,
because there is a small contribution from the Weinberg operator
proportional to the product $\kappa_t \tilde \kappa_t$. Our
numerical study shows that a factor 300 improvement in the measurement
of the neutron EDM will lead to ${\mathcal O}(10^{-3})$ constraints on
$\tilde \kappa_t$ from the Weinberg operator alone (and will thus not
dependent on assumptions about the Higgs couplings to the first 
generation fermions).  This sensitivity exceeds the projected precision 
of the Higgs-boson measurements at a high-luminosity upgrade of the LHC.

In the case of the Higgs-bottom and -tau couplings we find that the
present LHC Higgs data permit ${\cal O} (1)$ modifications in
$\kappa_{b,\tau}$ and $\tilde \kappa_{b,\tau}$.  While the EDMs
currently impose even weaker bounds, the situation may be
  reversed in the future. With $3000 \, {\rm fb}^{-1}$ of integrated
luminosity the LHC should be able to constrain ${\cal O} (0.05)$
values of $\kappa_b$, while the sensitivity of the proposed EDM
measurements reaches ${\cal O} (10^{-2})$ for $\tilde \kappa_b$, if the
Higgs boson couples to the first  generation with SM
strength. Assuming that the Higgs does only interact with the third
generation, extracting a limit from the neutron EDM on $\tilde
\kappa_b$ of ${\cal O} (0.05)$ should still be possible.  In the case
of the Higgs-tau couplings, we saw that even the full high-luminosity
LHC data set will allow for $\tilde \kappa_\tau = {\cal O}(1)$. 
Using angular correlations in the $h \to \tau \bar \tau$ channel, direct 
searches at the LHC may be capable  to probe $\tilde \kappa_\tau$ 
values of ${\cal O} (0.1)$~\cite{Berge:2008wi, Berge:2008dr,Berge:2011ij, 
Fermilab:CPV}. A possible improvement by three orders of magnitudes 
in the determination of the electron EDM will, on the other hand, make 
values  $\tilde \kappa_\tau = {\cal O} (10^{-2})$ accessible, if the Higgs 
couples to the electron.

\acknowledgments We are grateful to Junji Hisano and Koji Tsumura for
reminding us of the role of the threshold corrections to the Wilson
coefficient of the Weinberg operator and to Martin Jung for useful
correspondence concerning the mercury EDM. We would like to thank the
KITP in Santa Barbara, where this work was initiated, for warm
hospitality and acknowledge that this research was supported in part
by the National Science Foundation under Grant No. NSF
PHY11-25915. J.B. and J.Z. were supported in part by the U.S. National
Science Foundation under CAREER Grant PHY-1151392.
 
\appendix
\section{RG analysis for neutron EDM}
\label{app:neutronEDM:RG}

In order to estimate the size of the neutron EDM one has to perform a
RG analysis including the effects of operator mixing. The mixing of
the three operators
\begin{equation}
\begin{split}
{\cal Q}_1^q &= -\frac{i}{2} \hspace{0.25mm} e \, Q_q \, m_q \, \bar
q  \hspace{0.25mm}  \sigma^{\mu \nu} \gamma_5  \hspace{0.25mm}  q \,
F_{\mu \nu} \,, \\[1mm] 
{\cal Q}_2^q &= -\frac{i}{2} \hspace{0.25mm} g_s m_q \, \bar
q  \hspace{0.25mm}  \sigma^{\mu \nu} T^a \gamma_5  \hspace{0.25mm}  q
\, G^a_{\mu \nu} \,, \\[1mm] 
{\cal Q}_3 &= -\frac{1}{3}  \hspace{0.25mm} g_s  \hspace{0.25mm}
f^{abc} \, G_{\mu \sigma}^a G_{\nu}^{b, \sigma} \widetilde G^{c, \mu
  \nu} \,,  
\end{split}
\end{equation}
has been given in~\cite{Braaten:1990gq,Degrassi:2005zd}. In this normalization the
Wilson coefficients at the high scale read
\begin{equation}  \label{oldWC}
\begin{split} 
{\cal C}_1^q (\mu_W) & =  -\frac{16}{3}
\frac{\alpha}{(4\pi)^3} \hspace{0.25mm} \sqrt{2} G_F  \, \kappa_q
\tilde\kappa_t    \hspace{0.5mm} f_1 (x_{t/h}) \,, \\[1mm] 
{\cal C}_2^q (\mu_W) & = -2 \hspace{0.25mm} \,
\frac{\alpha_s}{(4\pi)^3} \hspace{0.25mm} \sqrt{2} G_F  \, \kappa_q
\tilde\kappa_t    \hspace{0.5mm} f_1 (x_{t/h}) \,, \\[1mm] 
{\cal C}_3 (\mu_W) & = \frac{1}{4} \frac{\alpha_s}{(4 \pi)^3} \,
\sqrt{2} G_F \, \kappa_t  \hspace{0.25mm} \tilde\kappa_t  \, f_3
(x_{t/h}) \,,  
\end{split}
\end{equation}
where $\alpha$ and $\alpha_s$ are understood to be evaluated at the
scale $\mu_W = {\cal O} (m_t)$. By solving the RG equations
\begin{equation} \label{eq:RGE}
\mu\hspace{0.5mm} \frac{d}{d \mu} \, \vec{{\cal C}} (\mu) = \gamma^T  \,  
\vec{{\cal C}} (\mu) \,, \qquad
\vec{{\cal C}} (\mu) = \big ({\cal{C}}_1^q (\mu), {\cal{C}}_2^q (\mu),
{\cal{C}}_3(\mu)\big  )^T \,,
\end{equation}
using the leading-order (LO) anomalous dimension matrix (ADM)
\begin{equation}
\gamma = \frac{\alpha_s}{4\pi} \begin{pmatrix} \frac{32}{3} &
  \phantom{-} 0 & 0 \\[2mm] 
\frac{32}{3} &\phantom{-} \frac{28}{3} & 0 \\[2mm] 
0 & -6  & 14 +  \frac{4 N_f}{3}  \end{pmatrix} \,,
\end{equation} 
with $N_f$ denoting the number of active flavors, one resums LL
effects and can determine the Wilson coefficients at the hadronic
scale $\mu_H$.

In~\cite{Pospelov:2005pr} a normalization of the three operators for
calculating their matrix elements is used that differs
from~\cite{Braaten:1990gq,Degrassi:2005zd}, namely
\begin{equation}
{\cal Q}_e^q = \frac{{\cal Q}_1^q}{e} \,, \qquad 
{\cal Q}_c^q = {\cal Q}_2^q \,, \qquad 
{\cal Q}_G =  \frac{{\cal Q}_3}{g_s} \,.
\end{equation}
In the latter basis, the Wilson coefficients at the high scale are given by 
\begin{equation} \label{newWC}
{\cal C}_e^q (\mu_W) = e \hspace{0.5mm} {\cal C}_1^q (\mu_W) \,, \qquad 
{\cal C}_c^q (\mu_W) = {\cal C}_2^q (\mu_W)  \,, \qquad 
{\cal C}_G (\mu_W) = g_s   (\mu_W) \hspace{0.5mm} {\cal C}_3 (\mu_W) \,. 
\end{equation}
The new Wilson coefficients (\ref{newWC}) can be obtained from the old
ones (\ref{oldWC}) by the simple redefinition $g_s(\mu_H) \, {\cal
  C}_3(\mu_W) = \eta^{-1/2} \, {\cal C}_G(\mu_W)$ with $\eta =
\alpha_s(\mu_W)/\alpha_s(\mu_H)$. Performing five- and four-flavor
running, we find in the new basis
\begin{equation}
\begin{split}
{\cal C}_e^q (\mu_H) & = 0.45 \, {\cal C}_e^q (\mu_W) - 0.38 \; {\cal C}_c^q 
(\mu_W)  - 0.07 \; \frac{{\cal C}_G (\mu_W)}{g_s (\mu_W)} \,,\\[1mm]
{\cal C}_c^q (\mu_H) & = 0.50 \, {\cal C}_c^q (\mu_W)  +0.15 \; 
\frac{{\cal C}_G (\mu_W)}{g_s (\mu_W)} \,, \\[3mm]
{\cal C}_G (\mu_H) & =  0.40 \, {\cal C}_G (\mu_W) \,.
\end{split}
\end{equation}
The above low-energy Wilson coefficients are related to the dipole
moments $d_q$, $\tilde d_q$, and the coefficient $w$ as follows
\begin{equation}
d_q = \hspace{0.25mm} Q_q \hspace{0.25mm} m_q \, C_e^q (\mu_H) \,, 
\qquad \tilde d_q = m_q \, C_c^q (\mu_H) \,, \qquad w = C_G (\mu_H) \,. 
\end{equation}
We now set $\mu_W = m_t$, $\mu_H = 1 \, {\rm GeV}$, and use $\alpha_s
(m_t) = 0.109$, $\alpha_s (\mu_H) = 0.36$, $m_u (\mu_H) = 2.4 \cdot
10^{-3} \, {\rm GeV}$, and $m_d (\mu_H) = 5.4 \cdot 10^{-3} \, {\rm
  GeV}$. These numerical values are obtained from the input given
in~\cite{Beringer:1900zz} and~\cite{Laiho:2011rb}, by employing
one-loop running of the strong coupling constant and the quark
masses. In this way, we arrive at
\begin{equation} \label{dnefinalsplit} 
\begin{split}
\frac{d_n}{e} & = \Big \{ (1.0 \pm 0.5) \left [ - ( 1.0\hspace{0.25mm}
  \kappa_u + 4.3 \hspace{0.25mm} \kappa_d )\hspace{0.5mm} \tilde
  \kappa_t + 5.1 \cdot
  10^{-2} \hspace{0.5mm} \kappa_t \tilde\kappa_t \right ] \\[3mm] &
   \phantom{xxx} + (22 \pm 10 ) \, 1.8 \cdot 10^{-2} \hspace{0.5mm} \kappa_t
\tilde\kappa_t \Big \} \cdot 10^{-25} \, {\rm cm} \,,
\end{split}
\end{equation}
where we kept the couplings to up and down quarks explicit. This
result generalizes the expression given in (\ref{dnefinal}).

\section{Bottom-quark contributions to neutron EDM}
\label{app:lightEDM:RG}

In Sec.~\ref{sec:lightEDMs} we have argued that integrating out the bottom 
quark together with the Higgs boson at the electroweak scale introduces a 
large scale uncertainty.  The situation can be remedied by removing the Higgs 
and the bottom quark as active degrees of freedoms in two steps and using 
RG-improved perturbation theory to  resum the logarithms in the 
expansions~(\ref{eq:dsw}) of the full results.

In a first step we integrate out the Higgs boson at the scale $\mu_W = {\cal O}
(M_h)$, which leads to an effective five-flavor theory. The corresponding 
Lagrangian is given in terms of the following operators 
\begin{equation} \label{eq:fullope}
\begin{split}
\Op_1^q & = \bar q q \,  \bar b  \hspace{0.25mm}  i \gamma_5  
\hspace{0.25mm}  b \,, \\[2mm]
\Op_2^q & = \bar q  \hspace{0.5mm}  T^a  \hspace{0.125mm}  q \, \bar b  
\hspace{0.25mm}   i \gamma_5 T^a   \hspace{0.25mm}  b \,, \\[2mm]
\Op_3^q & = \bar q  \hspace{0.25mm} \sigma_{\mu\nu}  \hspace{0.25mm}  
q \, \bar b  \hspace{0.25mm}  i \sigma^{\mu\nu}
  \gamma_5  \hspace{0.25mm}  b \,, \\[2mm]
\Op_4^q & = \bar q  \hspace{0.25mm}  \sigma_{\mu\nu} T^a  
\hspace{0.25mm}  q \, \bar b  \hspace{0.25mm}  i \sigma^{\mu\nu} \gamma_5
T^a  \hspace{0.25mm}  b \,, \\[2mm]
\Op_5^q & = -\frac{i}{2} \hspace{0.25mm} e \hspace{0.25mm} Q_b 
\hspace{0.25mm} \frac{m_b}{g_s^2} \,
\bar q \hspace{0.25mm} \sigma^{\mu \nu} \gamma_5 \hspace{0.25mm} 
q \hspace{0.5mm} F_{\mu \nu} \,,\\
\Op_6^q & = -\frac{i}{2} \hspace{0.25mm} \frac{m_b}{g_s} \, \bar q 
\hspace{0.25mm}
\sigma^{\mu \nu} T^a \gamma_5 \hspace{0.25mm} q \hspace{0.5mm} 
G^a_{\mu \nu} \,, \\
\Op_7 & = -\frac{1}{3 \hspace{0.125mm} g_s}   \hspace{0.25mm}  f^{abc}
\, G_{\mu \sigma}^a G_{\nu}^{b, \sigma} \widetilde G^{c, \mu \nu} \,,
\end{split}
\end{equation}
by 
\begin{equation} \label{eq:Leff}
\begin{split}
{\mathcal L}_\text{eff}   =  -\frac{\sqrt{2} G_F}{M_h^2}\,  m_b  \hspace{0.25mm} 
\tilde \kappa_b  \, \bigg \{  & \sum_{q=u,d}  m_q \hspace{0.25mm} \kappa_q  
\sum_{i=1,2,3,4,5,6} \Wc_i^q  \hspace{0.25mm} \Op_i^q  \\[2mm] & +   m_b  
\hspace{0.25mm} \kappa_b  \, \bigg [  \, \sum_{j=1,3,5,6} \Wc_j^b  \hspace{0.25mm} 
\Op_j^b + \Wc_7  \hspace{0.25mm} \Op_7 \, \bigg ] \bigg \} \,.
\end{split}
\end{equation}
Notice that the above operators are normalized in such a way that
operator mixing starts at $\mathcal{O} ( \alpha_s )$, $\gamma =
\alpha_s/(4 \pi) \, \gamma^{(0)} + (\alpha_s/(4 \pi))^2 \,
\gamma^{(1)} + {\cal O} (\alpha_s^3)$.  The Wilson coefficients
$\Wc_5^q$, $\Wc_6^q$, and~$\Wc_7$ are related to $d_q$, $\tilde d_q$,
and $w$ by 
\begin{equation} \label{eq:ddw}
\begin{split}
d_q (\mu) & = -\frac{e \hspace{0.25mm} Q_b}{4 \pi \alpha_s} \, \sqrt{2} G_F 
 \hspace{0.25mm}  m_q \, \kappa_q  \tilde\kappa_b  \,  x_{b/h}  \, \Wc_5^q (\mu)
  \,, \\[2mm]
\tilde d_q (\mu) & = -\frac{1}{4 \pi \alpha_s} \, \sqrt{2} G_F  \hspace{0.25mm}  
m_q \, \kappa_q  \tilde\kappa_b  \,  x_{b/h}  \, \Wc_6^q (\mu) \,, \\[2mm]
w (\mu) & = -\frac{g_s}{4 \pi \alpha_s} \, \sqrt{2} G_F  \hspace{0.25mm}
 \kappa_b  \tilde\kappa_b  \,  x_{b/h}  \, \Wc_7 (\mu) \,.
\end{split}
\end{equation}

\begin{figure}[!t]
\begin{center}
\vspace{-5mm}
\includegraphics[height=0.25\textwidth]{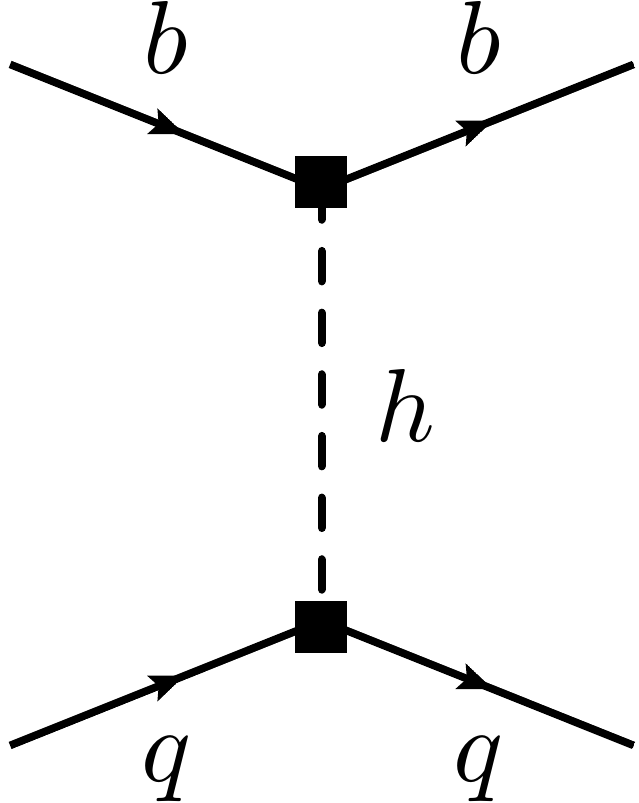} \qquad  \qquad 
\raisebox{2mm}{\includegraphics[height=0.225\textwidth]{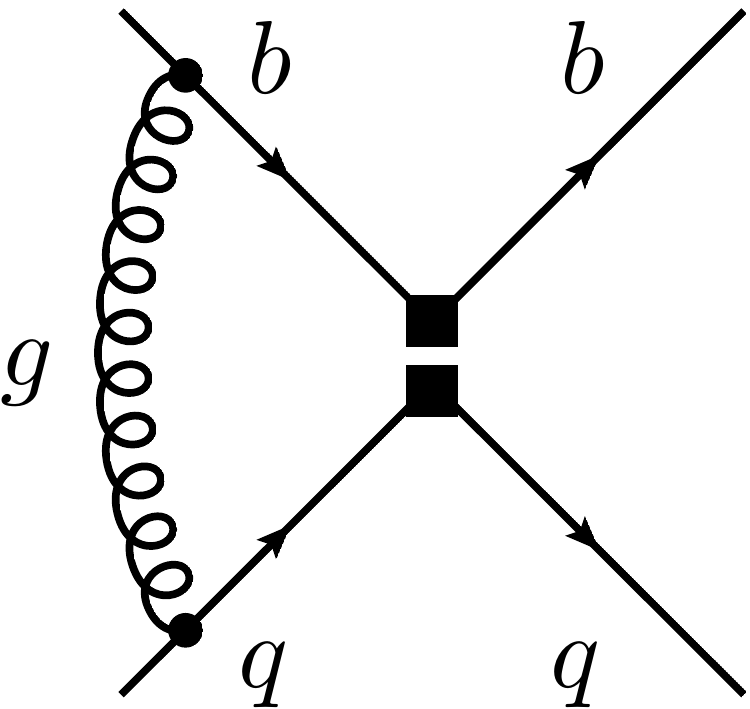}} 
\qquad  \qquad \includegraphics[height=0.265\textwidth]{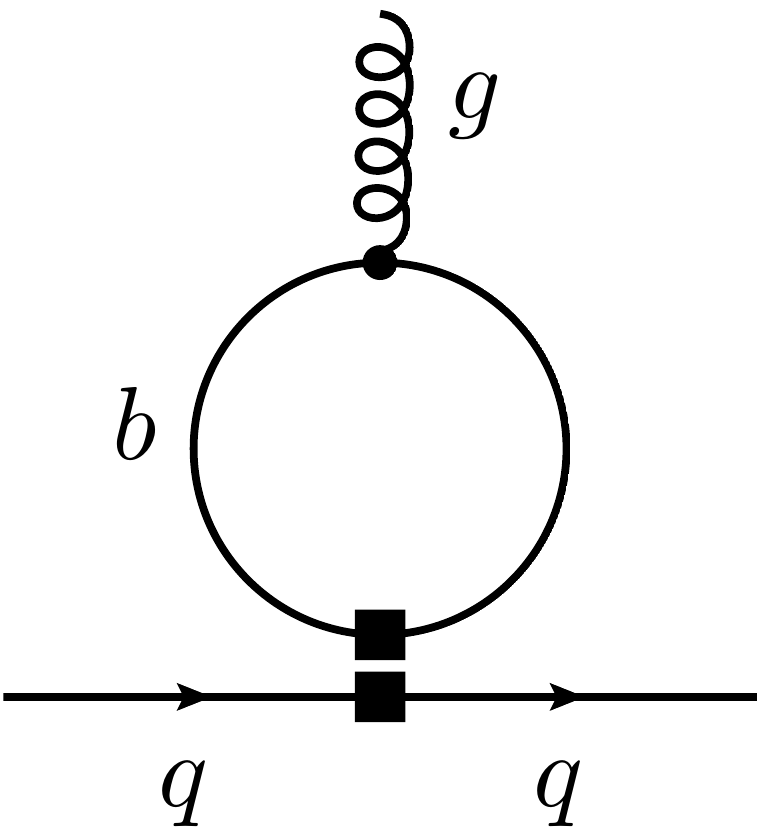}
\vspace{4mm}
\caption{\label{fig:matchb} {Left: The diagram in the full theory that gives rise
 to the LO matching correction to~$\Wc_1^q$. The same graph with the 
 quarks $q$ replaced by a bottom quarks $b$ induces a non-zero initial 
 condition for $\Wc_1^b$. Middle: An example of a graph that contributes 
 to the one-loop mixing of $\Op_1^q$ into $\Op_4^q$. Right:~Feynman 
 diagram describing the one-loop mixing of $\Op_4^q$ into $\Op_6^q$. 
 The shown graph involves a closed bottom-quark line. See text for further
  explanations.}}
\end{center}
\end{figure}

We perform the RG running between $\mu_W $ and the bottom-quark
threshold $\mu_b = {\cal O}(m_b)$ employing the operator basis
\begin{equation}
\vec \Op = (\Op_1^q, \Op_2^q, \Op_3^q, \Op_4^q, \Op_1^b, \Op_3^b,  
\Op_5^q, \Op_6^q,\Op_5^b, \Op_6^b, \Op_7)^T \,,   
\end{equation}
with $q = u,d$. At the tree level only  the Wilson coefficients of $\Op_1^q$ 
and $\Op_1^b$ receive a non-zero initial condition, cf.~Fig.~\ref{fig:matchb}~(left). 
In our normalization the corresponding matching coefficients  read 
\begin{equation}
\vec \Wc^{\, (0)} (\mu_W) =  (1, 0, 0, 0, 1, 0, 0, 0, 0, 0, 0)^T \,.
\end{equation}
Adapting existing results for the anomalous dimensions
\cite{Braaten:1990gq,Degrassi:2005zd,
  Borzumati:1999qt,Buras:2000if,Hisano:2012cc} to our definition of
operators~(\ref{eq:fullope}), we find for the LO ADM in the effective
five-flavor theory
\begin{equation} \label{eq:LOADM}
\gamma^{(0)} =
\begin{pmatrix}
-16&0&0&-2&0&0&0 & 0 &0 & 0 & 0 \\[2mm]
0&2&-\frac{4}{9}&-\frac{5}{6}&0&0&0 & 0 &0 & 0& 0  \\[2mm]
0&-96&\frac{16}{3}&0 &0 &0 &-48&0&0& 0 & 0\\[2mm]
-\frac{64}{3}&-40&0&-\frac{38}{3}&0 &0&0 &-8&0& 0& 0   \\[2mm]
0&0&0&0&-10&-\frac{1}{6}&0&0& 4& 4 & 0 \\[2mm]
0&0&0&0&40&\frac{34}{3}&0&0&-112& -16& 0  \\[2mm]
0&0&0&0&0&0&-\frac{14}{3}&0&0& 0 & 0 \\[2mm]
0&0&0&0&0&0&\frac{32}{3}&-6&0& 0 & 0 \\[2mm]
0 & 0 & 0 & 0 & 0 & 0 & 0 & 0 & -\frac{14}{3} & 0 & 0 \\[2mm]
0 & 0 & 0 & 0 & 0 & 0 & 0 & 0 & \frac{32}{3} & -6 & 0 \\[2mm]
0&0&0&0&0& 0&0&-6& 0& -6 & \frac{16}{3}
\end{pmatrix}\,. 
\end{equation}
Notice that the operators $\Op_1^b$ and $\Op_3^b$ only mix among
themselves at the one-loop level. This implies that they do not affect
the resummation of the LL QCD contributions to $C_5^q$ and $C_6^q$.
The one-loop mixing of $\Op_1^b$ and $\Op_3^b$ plays however an
important role in the calculation of the NLL corrections to $C_7$.

We first discuss the resummation of logarithms for the Wilson coefficient 
of $\Op_6^q$. By solving the usual RG equations $\big($see~Eq.~(\ref{eq:RGE})$\big)$,  
we obtain in terms of  $\eta_5 \equiv \alpha_s (\mu_W)/\alpha_s (\mu_b)$, the
following expression 
\begin{equation} \label{eq:C6qresum}
\begin{split}
\Wc_6^q (\mu_b) & = \frac{432}{2773 \hspace{0.5mm} \eta_5 ^{9/23}}+
\frac{0.07501}{\eta_5 ^{1.414}}+9.921 \cdot 10^{-4} \hspace{0.5mm}
\eta_5^{0.7184}- \frac{0.2670}{\eta_5 ^{0.6315}}+\frac{0.03516}{\eta_5
  ^{0.06417}} + {\cal O} (\alpha_s^3) \\[2mm] 
& \simeq \bigg(
\frac{\alpha_s}{4\pi} \bigg)^2 \, \frac{\gamma_{14}^{(0)}
  \gamma_{48}^{(0)}}{8} \, \ln^2 x_{b/h} + {\cal O} (\alpha_s^3) \,,
\end{split}
\end{equation} 
for the Wilson coefficient of the CEDM operator. Notice that the final
result only contains the LL correction which is proportional to the
combination $\gamma_{14}^{(0)}\gamma_{48}^{(0)}$ of one-loop anomalous
dimensions (cf.~\eqref{eq:LOADM}). Inserting (\ref{eq:C6qresum}) into
(\ref{eq:ddw}) we recover the result for $\tilde d_q$ as given
in~Eq.~(\ref{eq:dsw}).  Diagrammatically the LL bottom-quark
corrections to the CEDM therefore arise from the mixing $\Op_1^q \to
\Op_4^q$ followed by $\Op_4^q \to \Op_6^q$, cf.~Fig.~\ref{fig:matchb}
(middle and right). Employing $\mu_W = M_h$ and $\mu_b = m_b = 4.2 \,
   {\rm GeV}$ with $\alpha_s (M_h) = 0.113$ and $\alpha_s (m_b) =
   0.212$, we find numerically $\Wc_6^q (m_b) = 0.008 \; (0.026)$ for
   the last line in Eq.~(\ref{eq:C6qresum}) using $\alpha_s = \alpha_s
   (M_h)$ $\big($$\alpha_s = \alpha_s (m_b)$$\big)$. The resummed
   result is $\Wc_6^q (m_b) = 0.022$, which shows that the resummation
   of QCD logarithms is phenomenologically important for the CEDM
   $\tilde d_q$.

The Wilson coefficient of the operator $\Op_5^q$ receives both QED 
and QCD corrections.  Including the ${\cal O} (\alpha)$ contributions 
to $d_q$ as given in~Eq.~(\ref{eq:dsw}) in their unresummed form, 
but resumming the LL QCD effects, we obtain {
\begin{equation} \label{eq:C5q}
\begin{split}
\Wc_5^q (\mu_b) & = -4 \, \frac{\alpha \hspace{0.25mm}  \alpha_s}{(4 \pi)^2} \,
 Q_q \left ( \ln^2 x_{b/h} + \frac{\pi^2}{3} \right ) + \frac{2688}{2773 \hspace{0.5mm} \eta_5 ^{7/23}}-
\frac{3456}{2773 \hspace{0.5mm} \eta_5 ^{9/23}} \\[2mm] 
& \phantom{xx}  - \frac{0.03467}{\eta_5 ^{1.414}}+
0.01407\hspace{0.5mm} \eta_5^{0.7184}+\frac{0.4102}{\eta_5 ^{0.6315}}-
\frac{0.1126}{\eta_5 ^{0.06417}}  + {\cal O} (\alpha_s^4) \\[3mm] 
& \simeq -4 \, \frac{\alpha  \hspace{0.25mm}  \alpha_s}{(4 \pi)^2} \, Q_q  
\ln^2 x_{b/h}  + \bigg( \frac{\alpha_s}{4\pi} \bigg)^3 \,  
\frac{\gamma_{14}^{(0)}\gamma_{48}^{(0)}\gamma_{87}^{(0)}}{48} \, 
\ln^3 x_{b/h}  + {\cal O} (\alpha_s^4)  \,.
\end{split} 
\end{equation}    
One observes that the leading QCD contribution to $\Wc_5^q$ is of 
${\cal O} (\alpha_s^3 \hspace{0.25mm} \ln^3 x_{b/h} )$ and proportional 
to the product $\gamma_{14}^{(0)}\gamma_{48}^{(0)}\gamma_{87}^{(0)}$ 
of the elements of the LO ADM~\eqref{eq:LOADM}. The LL QCD effects are 
hence formally of  three-loop order, cf.~Fig.~\ref{fig:threeandtwo}~(left). It 
follows that the ratio between QCD and QED effects in $\Wc_5^q$ is 
approximately given by 
\begin{equation}
-\frac{2  \hspace{0.25mm}  \alpha_s^2}{9 \pi  \hspace{0.25mm}  
Q_q  \hspace{0.25mm}  \alpha} \,  \ln x_{b/h} \simeq \frac{3.0}{Q_q} \,.
\end{equation}
This shows that for  $q = u$ ($q = d$) QCD corrections dominate over 
the QED effects by a factor of around $4.5$ ($-9.0$).  In our numerical 
analysis we therefore employ the full result for the Wilson coefficient 
$\Wc_5^q$ as given in the first two lines of Eq.~(\ref{eq:C5q}). 
\begin{figure}[!t]
\begin{center}
\vspace{-5mm}
\includegraphics[height=0.265\textwidth]{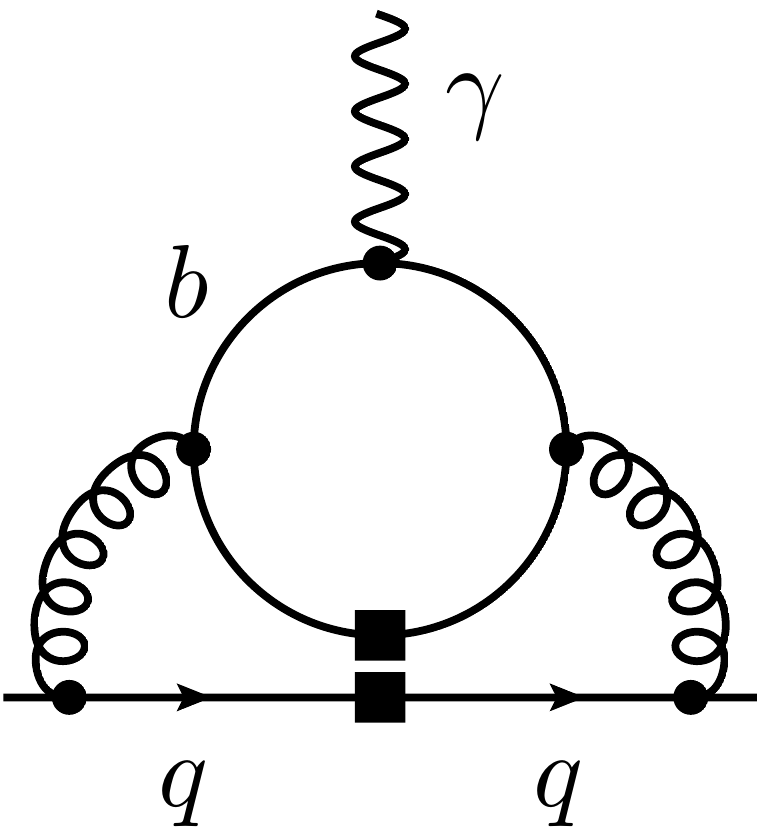} \qquad  \qquad \qquad 
\includegraphics[height=0.28\textwidth]{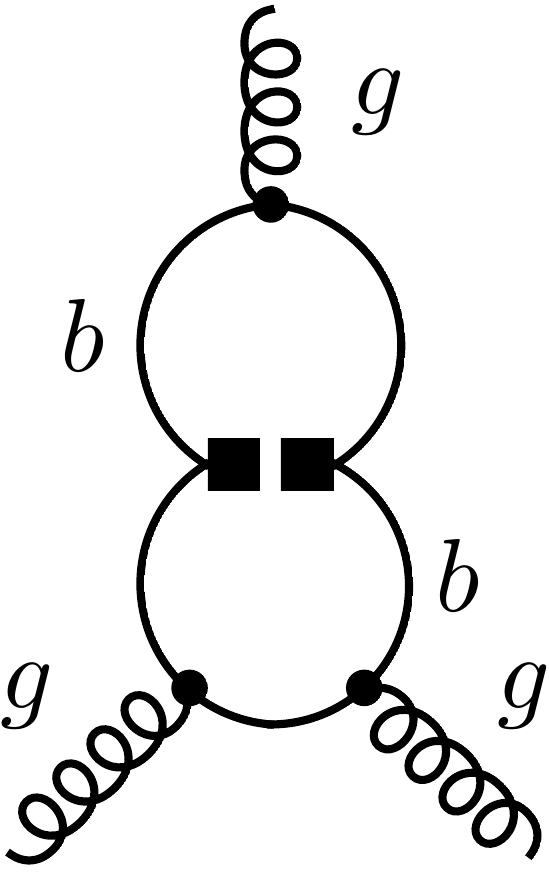}
\vspace{4mm}
\caption{\label{fig:threeandtwo}{Left: An example of a three-loop
    diagram involving an insertion of $\Op_1^q$ that gives rise to a
    logarithm of the form $\alpha_s^3 \ln^3 x_{b/h}$ in the Wilson
    coefficient $\Wc_5^q$. Right: A two-loop graph describing the
    mixing of $\Op_1^b$ into $\Op_7$.  For further details see text.}}
\end{center}
\end{figure}

In the case of the coefficient $w$ of the Weinberg operator the
resummation of the logarithmically-enhanced corrections
in~Eq.~(\ref{eq:dsw}) is slightly more involved as it requires the
knowledge of one- and two-loop anomalous dimensions. However, since
only the initial condition for $Q_1^b$ is non-vanishing at LO, the
only element needed from the ${\cal O} (\alpha_s^2)$ ADM to resum the
$\alpha_s \ln x_{b/h}$ term in $w$ is $\gamma_{5,11}^{(1)}$. This
element describes the mixing of the operator $\Op_1^b$ into $\Op_7$,
cf.~Fig.~\ref{fig:threeandtwo}~(right).  From the LL-expanded
expression for $w$, i.e.~Eq.~\eqref{eq:dsw}, we obtain
$\gamma_{5,11}^{(1)} = 2$.  To gain full control over the order
$\alpha_s^3 \ln^2 x_{b/h}$ terms in the Wilson coefficient $\Wc_7$
requires also the knowledge of the element $\gamma_{6,11}^{(1)}$ of
the two-loop ADM. By performing an explicit calculation we find that
$\gamma_{6,11}^{(1)} = 0$. Solving the RG equations then gives the
full two-loop result
\begin{equation} \label{eq:C7}
\begin{split}
\Wc_7 (\mu_b) & = \frac{\alpha_s (\mu_b)}{4 \pi } \left ( \frac{4200}{659}  \hspace{0.5mm} 
\eta_5^{31/23} + 0.3176 \hspace{0.5mm} \eta_5^{0.7184}  -\frac{6.691}{\eta_5 ^{0.6315}} \right ) 
\cdot 10^{-2} + {\cal O} (\alpha_s^3) \\[3mm] 
& \simeq   \bigg ( \frac{\alpha_s}{4\pi} \bigg)^2 \, 
 \frac{\gamma_{5,11}^{(1)}}{2} \, \ln x_{b/h}  + {\cal O} (\alpha_s^3)  \,.
\end{split} 
\end{equation} 
Using Eq.~(\ref{eq:ddw}), we recover from the final expression
the NLL contribution to $w$ as reported
in~Eq.~(\ref{eq:dsw}). Comparing the leading term in the expansion to
the resummed result, we find $\Wc_7 (m_b) = -0.6 \cdot 10^{-3} \; (-2.0
\cdot 10^{-3})$ for the last line in Eq.~(\ref{eq:C7}) employing
$\alpha_s = \alpha_s (M_h)$ $\big(\alpha_s = \alpha_s (m_b)\big)$,
while the RG-improved result is $\Wc_7 (m_b) = -1.2 \cdot 10^{-3}$. One
observes again that for an accurate description of the effects
associated to the Weinberg operator an RG analysis is mandatory.

Below the bottom-quark threshold one has to switch to the four-flavor
theory by integrating out the $b$ quark. We use the following reduced
set of operators \begin{equation} \vec \Op = \left ( \frac{m_q}{m_b} \, \Op_5^q,
\frac{m_q}{m_b} \, \Op_6^q, \Op_7 \right )^T\,, \end{equation} for which the
corresponding LO ADM reads~\cite{Degrassi:2005zd}
\begin{equation} \label{eq:LOADM4}
\gamma^{(0)} =
\begin{pmatrix}
-6&0&0\\[2mm]
\frac{32}{3}&-\frac{22}{3}&0\\[2mm]
0&-6&\frac{8}{3}
\end{pmatrix}\,. 
\end{equation}
The tree-level matching for the Wilson coefficients $\Wc_5^q (\mu_b)$
and $\Wc_6^q (\mu_b)$ is trivial, but at the one-loop level the CEDM
operator $\Op_6^b$ induces a finite threshold correction $\delta \Wc_7
(\mu_b)$ to the Wilson coefficient of the Weinberg operator when the
bottom quark is integrated out~\cite{Chang:1990jv}. The relevant
one-loop graphs are shown in Fig.~\ref{fig9}. We have
\begin{equation} \label{eq:dC7}
\delta \Wc_7 (\mu_b) =  \frac{\alpha_s (\mu_b)}{8 \pi} \, C_6^b (\mu_b) \,,
\end{equation}
with  
\begin{equation} \label{eq:C6bresum}
\Wc_6^b (\mu_b) = \frac{50}{47  \hspace{0.5mm} \eta_5^{9/23}} +  
3.969 \cdot 10^{-3} \hspace{0.5mm} \eta_5^{0.7184} 
-\frac{1.0678}{\eta_5^{0.6315}}  +{\cal O}  (\alpha_s^3) \,.
\end{equation}
Solving the RG equations, we then obtain for the Wilson
coefficients at the hadronic scale $\mu_H$ the following expressions 
\begin{equation}
\begin{split}
\Wc_5^q (\mu_H) & = \eta_4^{-9/25} \, \Wc_5^q (\mu_b) + 8 \left 
(  \eta_4^{-9/25} - \eta_4^{-11/25} \,  \right ) \Wc_6^q (\mu_b) \\[2mm] 
& \phantom{xx} +  \left (  \frac{72}{13} \, \eta_4^{-9/25} -  \frac{24}{5} \, 
\eta_4^{-11/25} -    \frac{48}{65} \, \eta_4^{4/25} \,  \right )  
\frac{\kappa_b}{\kappa_q} \, \Wc_7 (\mu_b) \,, \\[3mm]
\Wc_6^q (\mu_H) & = \eta_4^{-11/25} \,  \Wc_6^q (\mu_b) + \frac{3}{5}  
\left ( \eta_4^{-11/25} - \eta_4^{4/25} \,  \right )
  \frac{\kappa_b}{\kappa_q} \, \Wc_7 (\mu_b)  \,, \\[3mm]
\Wc_7 (\mu_H) & = \eta_4^{4/25}   \, \Big  (   \Wc_7 (\mu_b)  + 
 \delta \Wc_7 (\mu_b) \Big )   \,.
\end{split}
\end{equation} 
Here $\eta_4 \equiv \alpha_s (\mu_b)/\alpha_s (\mu_H)$ and the
results for $ \Wc_5^q (\mu_b)$, $ \Wc_6^q (\mu_b)$, $ \Wc_7
(\mu_b)$, and $\delta \Wc_7 (\mu_b)$ were given previously in Eqs.~(\ref{eq:C5q}),
(\ref{eq:C6qresum}), (\ref{eq:C7}), and (\ref{eq:dC7}). Notice that we have included 
the $\delta \Wc_7 (\mu_b)$ contribution only in the case of $\Wc_7 (\mu_H)$, since
it gives a $\alpha_s^3 \ln^2 x_{b/h}$ correction, which corresponds to a 
sub-leading logarithm in the case of $\Wc_{5}^q (\mu_H)$ and $\Wc_{6}^q (\mu_H)$. 

\begin{figure}[!t]
\begin{center}
\vspace{-5mm}
\includegraphics[height=0.275 \textwidth]{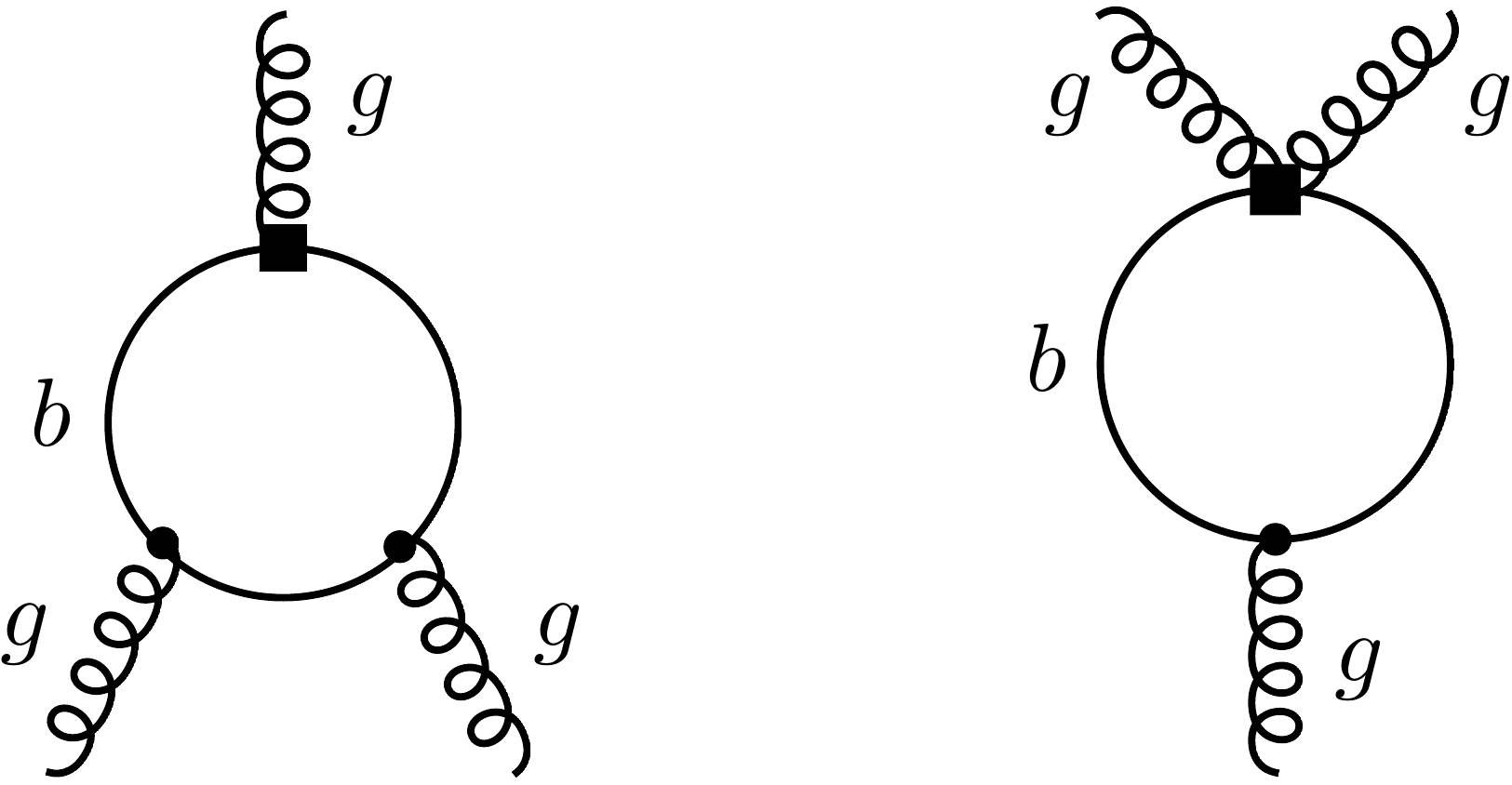} 
\vspace{4mm}
\caption{\label{fig9} One-loop diagrams leading to a correction to the Weinberg 
operator at the bottom-quark threshold. The black square denotes the insertion of the 
operator $\Op_6^b$. For further details consult the text.}
\end{center}
\end{figure}

\section{Other low-energy constraints}
\label{App:other}

In this appendix we consider constraints on the modifications of the
Higgs couplings to third-generation fermions that arise from
low-energy probes other than the EDMs. Although the constraints
discussed below turn out to be not very restrictive, we still mention
them for completeness.

We begin our survey in the quark-flavor sector. Naively one might
expect that the Higgs exchange in two-loop diagrams,
cf.~Fig.~\ref{fig10}~(left), would lead to a CP-violating contribution
to $B_s$--$\bar B_s$ mixing. Due to the symmetric nature of the diagrams,
however, there is no correction proportional to
$\kappa_t \hspace{0.25mm} \tilde\kappa_t $ at the level of
dimension-six operators (we checked this through an explicit
calculation). The contributions proportional to
$\kappa_t \hspace{0.25mm} \tilde\kappa_t $ do arise beyond dimension
six. As such they are suppressed by light quark masses and thus
unobservable in practice. In consequence, CP violation in $B_s$--$\bar
B_s$ mixing does not provide any relevant constraints on the
modifications of the Higgs-top couplings.

\begin{figure}[!t]
\begin{center}
\vspace{-5mm}
\raisebox{-4mm}{\includegraphics[height=0.24 \textwidth]{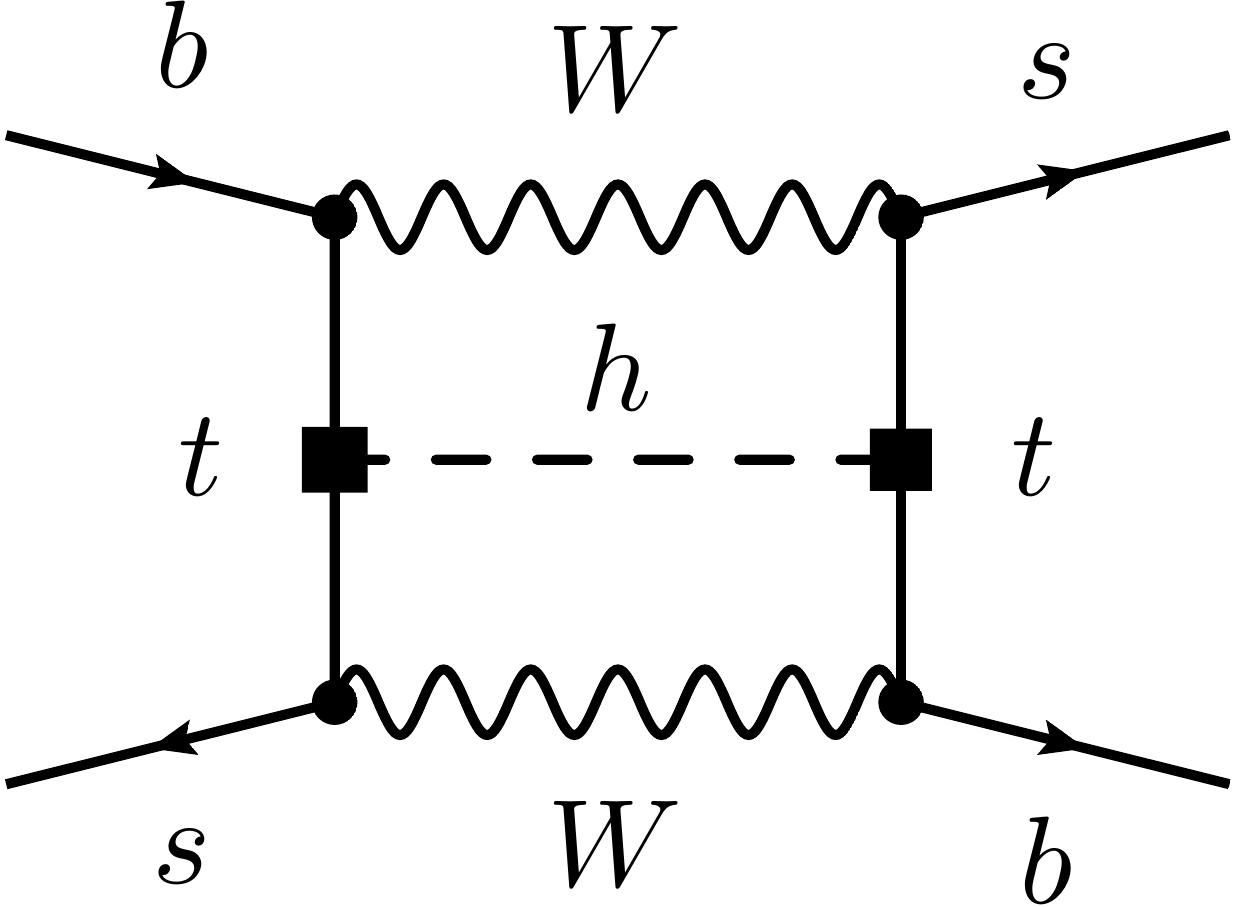}} \qquad \qquad
\includegraphics[height=0.2 \textwidth]{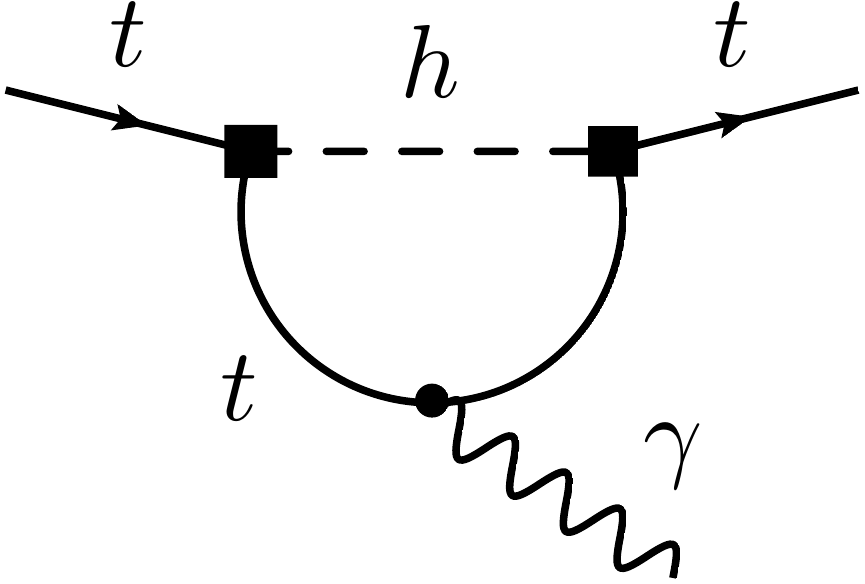} 
\vspace{4mm}
\caption{\label{fig10}  {Left: An example of a two-loop contribution to $B_s$--$\bar B_s$ 
mixing. Right: One-loop contribution to the magnetic dipole moment of the top quark. 
See text for details. }}
\end{center}
\end{figure}

Constraints on $\kappa_t$ and $\tilde\kappa_t $ in principle arise
also from the inclusive $B \to X_s \gamma$ decay at the two-loop
level. The resulting bounds can be estimated by calculating the
magnetic dipole moment of the top quark that is induced by Higgs
exchange as shown in~Fig.~\ref{fig10}~(right).  Inserting the effective
top-photon interaction into the two one-loop graphs in which the
photon is emitted from the internal top-quark line, i.e.~those with
$W$-boson and would-be Goldstone boson exchange, then leads to a
contribution to $b \to s \gamma$. Performing such an analysis one
finds that effects of ${\cal O} (100)$ in $\kappa_t $ and
$\tilde\kappa_t $ are not in conflict with $B \to X_s \gamma$. Notice
that an explicit two-loop calculation of $B \to X_s \gamma$
would~e.g.~also involve diagrams with $hWW$ vertices that are not
included in our estimate. However, we expect that a complete ${\cal O}
(\alpha^2)$ calculation would not change our general conclusions.

If the Higgs-fermion interactions are modified according to
Eq.~(\ref{higgsferm}) the one-loop amplitude of the $b \to s h$
transition is altered. Such a modification will change the prediction
for the branching ratio of $B_s \to \mu^+ \mu^-$. Allowing for a shift
in ${\rm Br} \left (B_s \to \mu^+ \mu^- \right )$ of $10^{-9}$ one can
derive that values of ${\cal O} (100)$ of both $\kappa_t $ and
$\tilde\kappa_t $ are consistent with the latter constraint.

In the case of $\kappa_\tau \neq 1$ and $\tilde \kappa_\tau \neq 0$,
one-loop Higgs exchange will modify the prediction for the anomalous
magnetic moment $a_\tau$ of the tau lepton. Modifications of ${\cal O}
(10)$ in the Higgs-tau couplings can be shown to result in shifts in
$a_\tau$ of ${\cal O} (10^{-8})$, i.e.~effects comparable to the
present SM uncertainty. Given that the experimental accuracy of the
measurements of $a_\tau$ are of ${\cal O} (10^{-2})$ even future
measurement of the anomalous magnetic moment of the tau are unlikely
to provide useful restrictions on $\kappa_\tau$ and $\tilde
\kappa_\tau$.

\end{document}